 \documentclass[final,5p,times]{elsarticle}

\usepackage{graphicx}

\usepackage{amssymb}
\usepackage{color}
\usepackage{psfrag,rotating}

\newcommand{\be}{\begin{equation}}
\newcommand{\ee}{\end{equation}}
\newcommand{\bea}{\begin{eqnarray}}
\newcommand{\eea}{\end{eqnarray}}

\newcommand{\vk}{\ensuremath{\vec{k}}}

\journal{New Astronomy}

\begin{document}

\begin{frontmatter}



\title{The mass of the dark matter particle: theory and galaxy observations}


\author[1,3]{H. J. de Vega\fnref{xx}}

\author[2]{P. Salucci}

\author[3]{N. G. Sanchez}

\address[1]{LPTHE, Universit\'e
Pierre et Marie Curie (Paris VI) et Denis Diderot (Paris VII),
Laboratoire Associ\'e au CNRS UMR 7589, Tour 13-14, 4\`eme. et 5\`eme. \'etage, 
Bo\^{\i}te 126, 4, Place Jussieu, 75252 Paris, Cedex 05, France}

\address[2]{SISSA/ISAS, via Beirut 4, I-34014, Trieste, Italia}

\address[3]{Observatoire de Paris, LERMA, Laboratoire
Associ\'e au CNRS UMR 8112,
61, Avenue de l'Observatoire, 75014 Paris, France}

\begin{abstract}
In order to determine as best as possible the nature of the dark matter (DM)
particle (mass and decoupling temperature) we compute analytically the DM galaxy properties 
as the halo density profile, halo radius and surface density and compare them to their 
observed values. We match the theoretically computed surface density to its observed value 
in order to obtain: 
(i) the decreasing of the phase-space density since equilibration till
today (ii) the mass of the dark matter particle and the decoupling temperature $ T_d
$ (iii) the kind of the halo density profile (core or cusp).
The dark matter particle mass turns to be between 1 and 2 keV and the
decoupling temperature $ T_d $ turns to be above 100 GeV.
keV dark matter particles necessarily produce cored density profiles while
wimps ($ m \sim 100 $ GeV, $ T_d \sim 5 $ GeV) inevitably produce
cusped profiles at scales about 0.003 pc. We compute in addition
the halo radius $ r_0  $, the halo central density
$ \rho_{0} $ and the halo particle r. m. s. velocity $ {\overline {v^2}}_{halo}^{1/2} $
they all reproduce the observed values within one order of magnitude.
These results are independent of the particle physics model and vary very little with 
the statistics of the dark matter particle. 
The framework presented here applies to any kind of DM particles: when applied to typical CDM  
GeV wimps, our results are in agreement with CDM simulations. keV scale DM particles 
reproduce all observed galaxy magnitudes within one order of magnitude
while GeV DM mass particles disagree with observations in up to eleven orders of magnitude.  

\end{abstract}


\fntext[xx]{Corresponding author}

\end{frontmatter}


\section{Introduction}

Since several years and more recently
\citep{disvdb,disvdb2,disvdb3,disvdb4} it has been  stressed
that basic galaxy parameters as mass, size, baryon-fraction, 
central density, 
are not independent from each other but in fact all of them do depend on
one parameter that works as a galaxy identifier. 
In fact there exist functional relations 
that constrain the different galaxy parameters in such a way that 
the galaxy structure depends essentially on one parameter (\citep{gts,sal07}
and references therein).

These functional relations may play for galaxies the r\^ole that
the equations of state play in thermodynamical systems.

First, let us remind that the density of DM in 
galaxies is usually well reproduced by dark halos 
with a cored distribution \citet{deb,sfm}, where 
$ r_0 $ is the core radius, $ \rho_0 $  is the central density 
$ {\displaystyle \lim_{r \to 0}} \; \rho(r)  = \rho_0 $ 
and $ \rho(r) $ for $ r<r_0 $ is approximately constant.
Recent findings highlight the quantity  
$ \mu_0 \equiv  r_0  \; \rho_0  $ 
proportional to the halo central surface density defined as   
$$ 
2 \int_0^{\infty}  \rho(0,0,x_3) \; dx_3 \quad {\rm where} \quad 
{\vec r}=(x_1,x_2,x_3) \; 
$$
where $ x_3 $ goes along the line of sight. The quantity 
$ \mu_0 $ is found nearly {\bf constant} and independent  of 
luminosity in  different galactic systems (spirals, dwarf irregular and 
spheroidals, elliptics) 
spanning over $14$ magnitudes in luminosity and  over different 
Hubble types. More precisely, all galaxies seem to have the same value 
for $ \mu_0 $, namely $ \mu_0 \simeq 120 \; M_\odot /{\rm pc}^2 $
\citep{kor,dona,span}.   It is remarkable that at the same time 
other important structural quantities as $ r_0 , \; \rho_0 $, 
the baryon-fraction and the galaxy mass vary orders of magnitude 
from one galaxy to another.

The constancy of $ \mu_0 $ is unlikely to be a coincidence and probably
has a deep physical meaning in the process of galaxy formation.
It must be stressed that $ \mu_0 $ is the only dimensionful quantity
which is constant among galaxies.

By analogy with the theory of phase transitions in statistical physics
we find useful to call 'universal' those quantities which 
take the same value for a large set of galaxies 
while non-universal quantities 
vary orders of magnitude from one galaxy to another. 
In this context the quantities
called universal take the same value up to $ \pm 20 \% $ 
for different galaxies.
 
Other known universal quantity in the above sense is the shape of the 
density profile when expressed as a function of $ r/r_0 $ 
and normalized to unit at $ r = 0 $. 

\medskip

In order to understand the above observations,
we compute here from the Boltzmann-Vlasov equation
\citep{dod,kt} the DM density profile and the surface density $ \mu_0 $
for different types of DM.
 
In this paper, we follow the evolution of the gravitational  
collapse of a perturbation of mass 
$ M \sim 3 \times 10^{12} M_{\odot} $ and derive  the 
 resulting linear halo density profile. This  reproduces  the 
phase of fast accretion found in $N$-body  simulations.  As a result, we
obtain robust predictions for the properties of DM halos.

In the case of  $\Lambda$CDM our results agree with 
the  $N$-body $\Lambda$CDM simulations and in the case of $\Lambda$WDM
our results agree with the observations. 

We obtain a very good fit of the computed profile to the Burkert 
profile. This determines the relation between $ r_0 $ and the free-streaming length.

\medskip

We also compute non-universal galaxy quantities as the halo radius, 
galaxy mass, halo central density and squared halo velocity. 
We find that the linear approximation
provides halo central densities smaller than or 
in the range of the observations, and halo velocities larger than 
the observed ones by a factor between 1 and 10.
We thoroughly analyze in our paper the validity of the linear approximation
to study galaxy properties and its limitations.
 Notice that
our determination of the DM particle mass does not relay to
these non-universal galaxy quantities.

\medskip

We combine the  observed properties of galaxies as
the effective core density and the core radius with the 
theoretical evolution of density fluctuations computed from first principles.

We consider in this paper the whole range of galaxy virial masses
going from 5 to 300 $ \times 10^{11} \; M_\odot $.

\medskip

\medskip

The theoretical treatment presented here captures many essential
features of dark matter, allowing to determine its nature.

Our treatment also applies to CDM: if we use the CDM surface density
value obtained from CDM simulations \cite{yh}, we determine [sec. \ref{conclu}] 
a dark matter particle mass in the wimps mass scale (GeV), fully consistent
with CDM simulations.

\medskip

This paper is organized as follows: Sec. 2 presents galaxy data and empirical
formulas relating basic galaxy parameters; sec. 3 deals with the phase-space density;
sec. 4 contains our theoretical results for the density profile from the linearized 
Boltzmann-Vlasov equation. In sec. 5 we derive the DM particle mass
and the decoupling temperature from the theoretical and observed 
galaxy surface density, in sec. 6 we compute non-universal galaxy properties
and in sec. 7 we derive the profiles for keV scale DM particles and for wimps
(cored vs. cusped profiles). In sec. 8 we present our conclusions.

\section{DM halos around galaxies: the observational framework}

The kinematics of about several thousands disk galaxies,  
described by the Universal Rotation Curves of Spirals, and the information
obtained from other tracers of the gravitational field of galaxies, 
including the  
dispersion velocities of spheroidals and the weak lensing measurements 
(\citep{gts,sal07} and references therein)
found that the density of the dark matter halos around galaxies of 
different kinds, different luminosity 
and Hubble types is well represented, out to the galaxy virial radius, 
by a Burkert profile
\be\label{bur}
\rho(r) =  \rho_0 \; F_B\left(\frac{r}{r_0}\right) \; , \;
F_B(x) =  \frac1{(1+x) \; (1 + x^2)} \; , \; x \equiv \frac{r}{r_0} \; ,
\ee
where  $ \rho_0 $ stands for the effective core density and $ r_0 $ for
the core radius. The Burkert profile satisfactorily fits the astronomical 
observations and we use the observed values of $ \rho_0 $ vs. $ r_0 $ for 
DM dominated spiral galaxies given in  \citep{sal07}.

\begin{figure}
\begin{turn}{-90}
\psfrag{"lmulr0.dat"}{$ \log_{10} r_0  $}
\includegraphics[height=9.cm,width=8.cm]{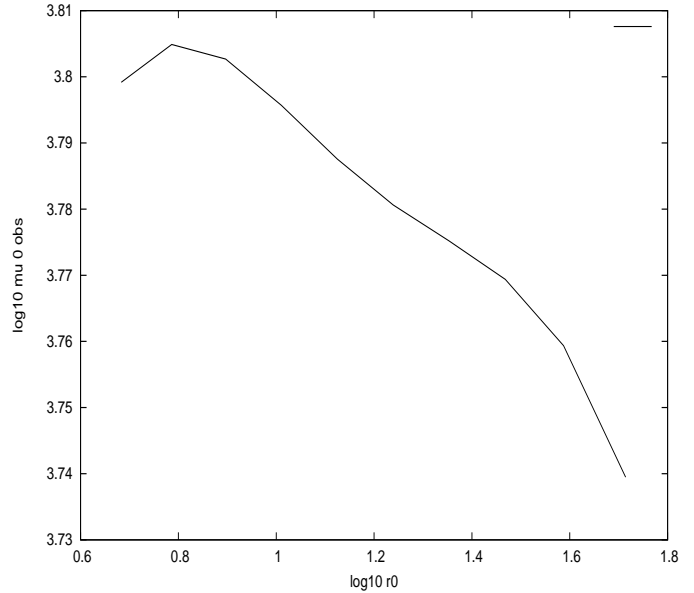}
\end{turn}
\caption{The common logarithm of the observed surface density  $ \mu_{0 \, obs} $ 
in $ ({\rm MeV})^3/ ( \hbar^2 \; c^4 ) $
vs. the common logarithm of the core radius $ r_0 $ in kpc.
Notice that in galaxies both $ r_0 $ and $ \rho_0 $ vary by a factor of
thousand while $ \mu_0 $ varies only by about $ \pm 20 \% $.}
\label{muo}
\end{figure}

\begin{table*}
  \begin{tabular}{cc} \hline  
  $ r_0 $ (kpc) & $  \mu_{0 \, obs} $ (MeV$^3$)  \\ \hline  \hline 
 4.8         &   0.63 $ \; 10^4 $ \\ \hline
 6.1 &   0.64 $ \; 10^4 $ \\ \hline
 7.9 &   0.63 $ \; 10^4 $  \\ \hline
 10.2 &   0.62 $ \; 10^4 $ \\ \hline
 13.3 &   0.61 $ \; 10^4 $ \\ \hline
 17.3 &   0.60 $ \; 10^4 $ \\ \hline
 22.6 &   0.60 $ \; 10^4 $ \\ \hline
 29.4 &   0.59 $\;  10^4 $ \\ \hline
 38.7 &   0.57 $\;  10^4 $ \\ \hline
51.8 &    0.55 $ \; 10^4 $  \\ \hline 
\end{tabular}
\caption{The observed core radius $ r_0 $ and the 
observed surface density  $ \mu_{0 \, obs} $ .}
\end{table*}

The structural halo parameters $ \rho_0 $ and $ r_0 $ are found to be 
related, it is worth to compute from them the virial mass $ M_{vir} $  
in terms of the core radius $ r_0 $ (\citep{gts,sal07} and references therein)
\be\label{mvir}
m_{\rm v} \equiv \frac{M_{vir}}{10^{11} M_{\odot}} = 0.320 \; 
\left(\frac{r_0}{\rm kpc}\right)^{1.72}
\ee
The surface density $ \mu_0 $ is defined as:
\be\label{denss}
\mu_0 \equiv \rho_0 \; r_0
\ee
We display in Table 1 the values of  the observed surface density  
$ \mu_{0 \, obs} $ in 
$ ({\rm MeV})^3 / ( \hbar^2 \; c^4 ) $ and the corresponding core radius 
$ r_0 $.
We plot in fig. \ref{muo} the observed surface density  $ \mu_{0 \, obs} $ in 
$ ({\rm MeV})^3 / ( \hbar^2 \; c^4 ) $ vs. the core radius $ r_0 $.

Notice that in galaxies  both $ r_0 $ and  $ \rho_0 $ vary by a factor 
$ 10^3 $ while $ \mu_0 $ varies only by less than $ \pm 20 \% $. 
$ 5 \; {\rm kpc} \lesssim  r_0 \lesssim 50 \; {\rm kpc} $
for normal spiral galaxies. Therefore, as stressed by \citep{kor,dona,span}
the surface density is a constant over a large number of 
galaxies of different kinds.

\medskip

Notice that the surface density of ordinary matter in luminous galaxies 
is about a factor 4 larger than the surface density value for
dark matter \cite{gg}. Clusters of galaxies, exhibit a dark matter surface density
about a factor 4 or 5 times larger than that of dwarf, elliptical and
spiral galaxies \cite{bs,et}. Such difference could be due to a baryons effect, 
which study is beyond the scope of this paper. For clusters of galaxies $ r_0 $ 
is 4 to 50 times larger than for the galaxies in Table 1 
and the masses are 100 to 4000 larger than the masses of the galaxies in Table 1. 
Namely, the variation of $ \mu_0 $ from galaxies to clusters of galaxies is a much 
small factor than the change in $ r_0 $ and in the total mass. 
We choose for the present work the data from galaxies in Table 1 (further discussion
on clusters of galaxies is given in sec. 6).  

\section{The invariant phase-space density of DM galaxy halos}\label{denesp}

The invariant phase-space density is defined by \citep{dm1,dep2,dm2,dep}
\be\label{defQ}
Q \equiv \frac{\rho}{\sigma^3} \qquad {\rm where}  \qquad 
\sigma^2 \equiv \frac13 \; <v^2> 
\ee
is the velocity dispersion. $ Q $ is invariant under the expansion of the 
universe and decreases due to self-gravity interactions \citep{thlb}
from its primordial value $ Q_p $ to the volume average value 
$ Q_{halo} $ of the galaxies today:
\be\label{defZ}
Q_{halo} = \frac1{Z} \; Q_p\; , 
\ee
where
\be\label{defZ2}
Q_{halo} \equiv \frac{\rho_{halo}}{\sigma^3_{halo}} \quad , \quad
Q_p\equiv \frac{\rho_{prim}}{\sigma^3_{prim}} \; .
\ee
This equation defines the factor $ Z $ \citep{dm2}. $ Z $ 
is larger than unity and its value depends on the galaxy considered.

Let us anticipate that $ Q_p $ only depends on the properties of the DM particle
and its primordial distribution function [see eq.(\ref{Qprim}) below]. 

\medskip

The velocity $ v_{halo}(r) $ follows from the virial 
theorem combined 
with the Burkert profile eq.(\ref{bur}) \citep{gts,sal07}
\bea\label{vbur}
\hskip -1cm v_{halo}^2(r) = 2 \, \pi \; G \; \frac{\rho_0 \; r_0^3}{r} \; 
\left[ \ln(1+x) -\arctan x + \frac12 \; \ln(1+x^2) \right] \; ,  &&\cr \cr
\hskip -9truecm x = \frac{r}{r_0}  \qquad \; . \qquad &&
\eea
$ Q_{halo} $ is obtained  by averaging $ \rho(r) $ and $ v^2_{halo}(r) $
over the volume using the density itself $ \rho(r) $ as weight function
(see \ref{averps}). From eqs. (\ref{bur}), (\ref{defQ}), 
(\ref{defZ2}) and (\ref{vbur}) we obtain [see eq.(\ref{vhprom})],
\be\label{Qpro}
 Q_{halo} = \frac{0.069}{G^\frac32 \; \sqrt{\rho_0} \; r_0^3} 
\quad {\rm (Burkert)} \; .
\ee
For a NFW profile,
\be\label{nfw}
\rho(r) =\frac{\rho_s}{\displaystyle \frac{r}{r_s} \displaystyle
\left(1 + \frac{r}{r_s} \right)^2} \; ,
\ee
we get:  
\be\label{Qnfw}
Q_{halo} = \frac{0.324}{G^\frac32 \; \sqrt{\rho_s} \; r_s^3} \quad {\rm (NFW)} \; .
\ee
Both results eqs. (\ref{Qpro}) and (\ref{Qnfw}) are of the same
order of magnitude and differ by a factor $ \sim 5 $.
Since $ Q \sim m^4 $ as shown below in eq.(\ref{Qprim}), using the cuspy
NFW profile instead of the cored Burkert profile only may change
the DM particle mass by a factor $ \sim 1.5 $ keeping its  order
of magnitude.

\begin{figure}
\begin{turn}{-90}
\psfrag{"QG11ufa.dat"}{$ \log_{10} Q^{-1}_{halo} $ vs. $ m_{\rm v} $}
\includegraphics[height=9.cm,width=9.cm]{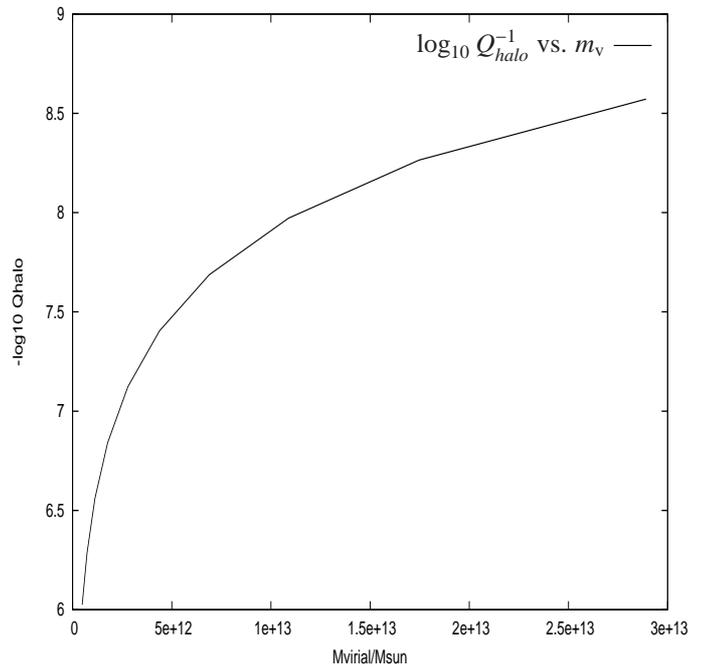}
\end{turn}
\caption{The logarithm${}_{10} $ of the phase-space density $ Q_{halo} $ 
obtained from eq.(\ref{Qpro}) using the data in Table 1
vs. the virial mass of the galaxy $ M_{virial} $ in units of solar masses $ M_\odot $.}
\label{Q}
\end{figure}

We plot in fig. \ref{Q} the phase-space density $ Q_{halo} $
vs. the virial mass of the galaxy $ M_{virial} $ in units of solar masses $ M_\odot $.
Notice that the virial mass of the galaxy 
is related to the halo radius $ r_0 $ through eq.(\ref{mvir}).

The primordial invariant phase-space density $ Q_p $ can be 
evaluated in the radiation 
dominated (RD) era with the result \citep{dm2}
\be\label{Qprim}
Q_p = \frac{3 \, \sqrt3}{2 \; \pi^2} \; g \; 
\frac{I_2^{\frac52}}{I_4^{\frac32}} \;  \frac{m^4}{\hbar^3} \; , 
\ee
where $ I_2 $ and $ I_4 $ are the dimensionless
momenta of the particle DM primordial distribution function \cite{dm2}:
$$
I_2 \equiv \int_0^{\infty} y^2 \; F_d(y) \; dy \; , \;
I_4 \equiv \int_0^\infty y^4  \; F_d(y)  \; dy \; ,
$$
$ g $ is the number of internal degrees of freedom of the DM 
particle ($ g = 2 $ for Dirac fermions). 
For example, for Dirac fermions of mass $ m $
that decoupled ultrarelativistically at thermal equilibrium we have,
\be\label{002}
Q_p = 0.020395 \; \frac{m^4}{\hbar^3} \; .
\ee
Similar expressions and values are obtained for bosons and for particles 
decoupling 
ultrarelativistically out of thermal equilibrium \citep{dm2}.

\medskip

The covariant decoupling temperature $ T_d $ can be expressed
in terms of the number of ultrarelativistic degrees of freedom at 
decoupling $ g_d $ by using entropy conservation \citep{gb} 
\be\label{temp}
T_d = \left(\frac2{g_d}\right)^\frac13 \; T_{\gamma} \; . 
\ee
$ g_d $ can be expressed as \cite{dm2}
\be\label{gdu}
g_d = \frac{2^\frac14}{3^\frac38 \; \pi^\frac32} \; 
\frac{g^\frac34}{\Omega_{DM}} \; \frac{T_{\gamma}^3}{\rho_c} \; 
Q_p^\frac14 \; \left(I_2 \; I_4\right)^{\frac38}
\ee

where $ T_{\gamma} $ is the CMB temperature today, $ \Omega_{DM} $ the DM cosmological 
fraction and  $ \rho_c $ the critical density of the universe.
From WMAP/LSS data we have \cite{pdg},
\bea\label{val0}
&& T_{\gamma} = 0.2348 \; {\rm meV } \; , \; \Omega_{DM} = 0.228\; , \cr \cr
&& \rho_c = (2.518 \; {\rm meV})^4 /(\hbar^3 \; c^5) \; ,
\eea
here 1 meV = $ 10^{-3} $ eV. 
We have in addition \cite{dm2}, 
\be\label{masa}
m = \pi^2 \; \Omega_{DM} \; \frac{\rho_c}{T_{\gamma}^3} \;
\frac{g_d}{g \; I_2}=
6.986 \; \mathrm{eV} \; \frac{g_d}{g \; I_2} \; , 
\ee
Hence, a DM particle decoupling ultrarelativisticaly at redshift $ z_d $ 
and physical decoupling temperature
$ T_d^{phys} = (1+z_d) \;  T_d \gtrsim 100 $ GeV where
$ g_d \sim 200 $ (see ref. \cite{kt}) will have a mass in the keV scale.

\section{The linear 
Boltzmann-Vlasov equation.}\label{perfil}

We now evolve the density fluctuations from the end of inflation till today
in the standard model of the Universe.
This evolution provides the phase-space density $ Q_{halo} $ and the surface density 
$ \mu_0 $ today. The density fluctuations follow from the distribution function
which evolves according to the non-linear Boltzmann-Vlasov equation.
The evolution is practically linear in the RD era and in the MD era before
structure formation. That is, we can use the linear Boltzmann-Vlasov for
redshift $ z \gtrsim 30 $. For  $ z \lesssim 30 $ non-linearities
are relevant and one should use the non-linear Boltzmann-Vlasov equation
or, alternatively, perform $N$-body simulations.
It must be noticed that the resolution of the linearized Boltzmann-Vlasov 
equation from the end of inflation till today provides a good approximated 
picture of the structures today \citep{ds1}. 
From the evolution of the dark matter fluctuations $ \Delta(k,z) $ 
we obtained the density profile 
$ \rho_{lin}(r) $  \citep{ds1}.

We follow the density fluctuations 
in the RD era
according to the results in \cite{dod} and \cite{husu}. 
It is convenient to recast
the linearized Boltzmann-Vlasov equation in the matter dominated 
(MD) era as an integral equation, the Gilbert equation \citep{gbs}. 
We solve  the Gilbert equation  \citep{ds1,gil}
to obtain the density fluctuations $ \Delta(k,z) $ till today 
\be\label{flueq}
\Delta(k,z)  \buildrel{z \to 0}\over= \frac35 \; T(k) \; (1 + z_{eq}) \; 
\Delta(k,z_{eq}) \; .
\ee
Here the subindex $_{eq}$ refers to equilibration, 
the beginning of the MD era, $ 1 + z_{eq} \simeq 3200 $
and $ T(k) $ is the transfer function which takes into account the
evolution of the density fluctuations during the matter dominated era.
$ T(k) $ has the properties:  $ T(0) = 1 $ and $ T(k \to \infty) = 0 $. 
Namely, the transfer function $ T(k) $ suppresses the large $ k $ 
(small scale) modes. 

It is convenient to introduce the dimensionless variable 
\be\label{defga}
\gamma \equiv k \; r_{lin}  \quad {\rm where} \quad   
r_{lin} \equiv \frac{l_{fs}}{\sqrt3} = \frac{\sqrt2}{k_{fs}} \; ,
\ee
$ l_{fs} $ and $ k_{fs} $ stand for the free-streaming length and 
free-streaming wavenumber respectively \cite{kt} and $ r_{lin} $
is given by \cite{gil}
\be\label{rlin1}
r_{lin} = 2 \; \sqrt{1 + z_{eq}} \;
\left(\frac{3 \; M_{Pl}^2}{H_0 \; \sqrt{\Omega_{DM}} \; Q_p}\right)^\frac13 \; ,
\ee
$ H_0 $ stands for the Hubble constant today and $ M_{Pl} $ for the Planck mass,
\be\label{val1}
H_0 = 1.5 \; 10^{-33} \; {\rm eV} \; , \;
M_{Pl} = 2.43 \; 10^{18} \; {\rm GeV} \; .
\ee
$ r_{lin} $ is the characteristic length scale in the linear regime.

We plot in fig. \ref{tk} the transfer function $ T(\gamma) $ for Fermions (FD) and Bosons 
(BE) decoupling ultrarelativistically, and for particles decoupling non-relativistically 
[Maxwell-Boltzmann statistics, (MB)]. We see from fig. \ref{tk} that the transfer function 
$ T(\gamma) $ decreases by an amount of order one for $ \gamma $ increasing by unit.
Therefore, $ T(k) $ decreases by an amount of order one when $ k $ increases
by an amount of the order of the wavenumber $ k_{fs} $ [see eq.(\ref{defga})].
As we see from fig. \ref{tk}, $ T(\gamma) $ shows little variation with the statistics
of the DM particles.

\subsection{The phase density from the
 observed and theoretical  surface density}\label{calcmu}

We match in this section the observed surface density (Table 1) with
the surface density computed from eqs. (\ref{muteo}) and (\ref{muteo2}).
This gives as a result eq.(\ref{eqq}) which
determines the primordial phase density.

We first compute the linearized density profile from the 
Fourier transform of the density fluctuations today
\citep{ds1}
\be\label{defro}
\rho_{lin}(r) = \frac1{2 \, \pi^2 \; r} \; \int_0^{\infty} k \; 
dk \; \sin(k \, r) \; \Delta(k,z = 0) \; ,
\ee

\begin{figure}
\begin{turn}{-90}
\psfrag{"tkfdgiR.dat"}{Fermi-Dirac}
\psfrag{"tkbegiR.dat"}{Bose-Einstein}
\psfrag{"r2tkmbgiR.dat"}{Maxwell-Boltzmann}
\includegraphics[height=9.cm,width=9.cm]{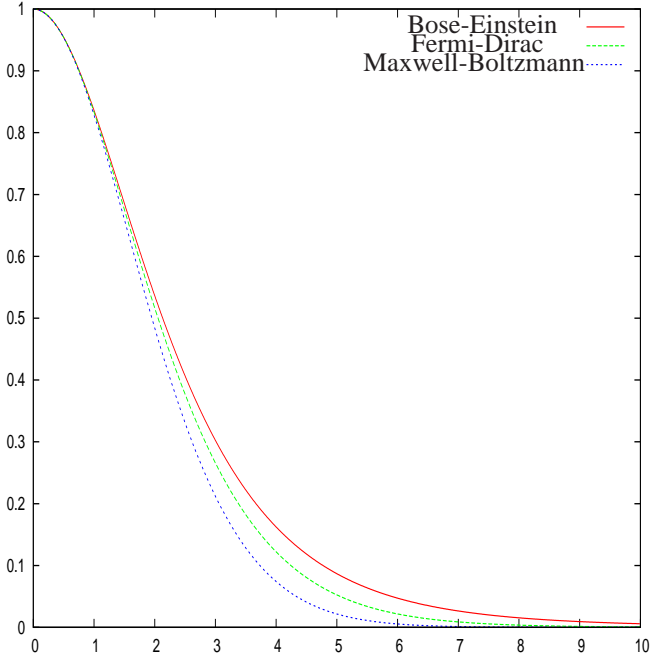}
\end{turn}
\caption{The transfer function $ T(k) $ vs. $ \gamma = k \; r_{lin} $
for Fermions and Bosons decoupling ultrarelativistically 
and for particles decoupling 
non-relativistically (Maxwell-Boltzmann statistics). 
$ T(\gamma) $ shows little variation with the statistics of the DM particles.
We see that $ T(k) $ decays for increasing $ k $ with a characteristic scale 
$ \sim  1/r_{lin} \sim k_{fs} $ which is the free-streaming wavenumber 
[see eq.(\ref{defga})].}
\label{tk}
\end{figure}

More explicitly, from eq.(\ref{flueq}) the density profile $ \rho_{lin}(r) $ 
turns to be the Fourier transform 
of the density fluctuations $ \Delta(k,z_{eq}) $ by the end of the RD era
times the transfer function $ T(k) $:
\bea\label{perf}
&&\rho_{lin}(r) = 
\frac{108 \; \sqrt2}{5 \; \pi} \; \frac{\Omega_{DM} \; M_{Pl}^2}{H_0}
\; (1+z_{eq}) \; |\Delta_0|  \cr \cr
&& \times   b_0 \; b_1 \;
\frac{k_0^{2-n_s/2}}{r_{lin}^{n_s/2} \; r} 
\int_0^{\infty} d\gamma \; N(\gamma) \; 
\sin\left(\gamma \, \frac{r}{r_{lin}} \right) \; ,
\eea
where $ | \Delta_0 | $ stands for the primordial power amplitude, 
$ n_s $ is the
primordial spectral index, $ k_0 $ is the pivot wavenumber 
used by WMAP to fit the primordial power, $ k_{eq} $ the horizon 
wavenumber by equilibration and
\be\label{defN}
N(\gamma) \equiv \gamma^{n_s/2-1} \; 
\ln\left(\frac{c_0 \; \gamma}{k_{eq} \; 
r_{lin}}\right) \; T(\gamma) \; .
\ee
The numerical values of the cosmological parameters
entering in eq.(\ref{perf}) are \cite{WMAP}
\bea\label{val2}
&&|\Delta_0 | \simeq 4.94 \quad 10^{-5} 
\quad , \quad n_s \simeq 0.964 \quad , \quad k_0 = 2 \; {\rm Gpc}^{-1} 
\; , \cr \cr
&& k_{eq}= 9.88 \; {\rm Gpc}^{-1} \quad , \quad c_0 \simeq 0.1160 \; .
\eea 
All fluctuations with $ k > k_{eq} $ that were inside the horizon by 
equilibration are relevant here \citep{ds1}. This introduces in 
eq.(\ref{perf}) the comoving horizon volume by equilibration \citep{ds1,dod}
\be\label{V}
\frac{b_1}{k_{eq}^\frac32}  \simeq \frac{ b_1 \; b_0}{H_0^\frac32} \; ,
\ee
where $ b_0  \simeq  3.669 \; 10^{-3} $ and
$ b_1 \sim 1 $ (actually, $ b_1 = 1 $ in \citep{ds1}).

The initial power fluctuations are multiplied by a Gaussian random field $ g(\vk) $
with unit variance
\be\label{gg}
< g(\vk) \; g^*(\vk') > = \delta(\vk-\vk') \; ,
\ee
which describes the random quantum character of the primordial fluctuations.

Each realization of the random field $ g(\vk) $ with unit variance and zero average
produces a DM configuration in the linear regime (a `galaxy').
The simplest one is obtained for $ g(\vk) = 1 $. 
The presence of $ g(\vk) $ will produce a large variety of non-spherically
symmetric galaxy configurations in a large range of masses and sizes. 
For simplicity we restrict ourselves here to the case $ g(\vk) = 1 $
and leave the inclusion of $ g(\vk) \neq 1 $ to future work. 
The profile $ \rho_{lin}(r) $ [with  $ g({\vec k}) = 1 $] 
bears the universal properties of the galaxies, that is to say, 
the general properties common to all (or most) galaxies.
This is why such a profile is very appropriate and useful to extract these 
universal properties.

\medskip

From the results eqs.(\ref{perf})-(\ref{val2}) we compute and analyze the 
surface density and the density profile.
We see from  eq.(\ref{perf}) that $ \rho_{lin}(r) $ decreases 
with $ r $ having $ r_{lin} $ as characteristic scale since it 
depends on $ r/r_{lin} $ being the Fourier
transform of a function of $ \gamma $ that decreases with unit characteristic
scale in $ \gamma $ [see fig. \ref{tk}]. 

We plot in fig. \ref{3perf} the ratio
\be\label{perfun}
\hskip -.6cm \frac{\rho_{lin}(r)}{\rho_{lin}(0)} \equiv \Psi(y) =
\frac{\int_0^{\infty} N(\gamma) \; \sin\left(\gamma \, y \right) \; 
d\gamma}{y \; \int_0^{\infty} \; \gamma \; 
N(\gamma) \; d\gamma} \; {\rm where} \; 
y \equiv \frac{r}{r_{lin}} \; ,
\ee
for Fermions (FD) and Bosons (BE) decoupling 
ultrarelativistically and for particles decoupling non-relativistically 
[Maxwell-Boltzmann statistics (MB)]. 

$ \Psi(y) $ mainly depends
on known cosmological parameters and fundamental constants and has a weak logarithmmic
dependence on the DM particle mass.

We compute theoretically the surface density from the density
profile eq.(\ref{perf}) and the halo radius  eqs.(\ref{rlin1}) and (\ref{r0rlin}).
Then,
\be\label{dfmulin}
\mu_{0 \, lin} \equiv r_0 \; \rho_{lin}(0) \; ,
\ee
with eqs. (\ref{rlin1})-(\ref{perf}) and (\ref{r0rlin}), $ \mu_{0 \, lin} $ reads: 
\bea\label{muteo}
&&\mu_{0 \, lin} = \frac{108 \, \sqrt2}{5 \, \pi} \; \Omega_{DM} \; 
 |\Delta_0 | \; 
(1 + z_{eq})^{1-n_s/4} \;  \frac{k_0^2 \; M_{Pl}^2}{H_0 \; \alpha} \cr \cr
&& \times   b_0 \; b_1 \; 
\left(\frac{\sqrt{\Omega_{DM}} \; H_0 \; Q_p}{24 \; k_0^3 \;  
M_{Pl}^2}\right)^{n_s/6}
\; \int_0^{\infty} \gamma \; N(\gamma) \; d\gamma \; .
\eea

\begin{figure}
\begin{turn}{-90}
\psfrag{"lqmv.dat"}{$ \log_{10} q \; $  vs. $ \log_{10} m_{\rm v} $}
\includegraphics[height=9.cm,width=7.cm]{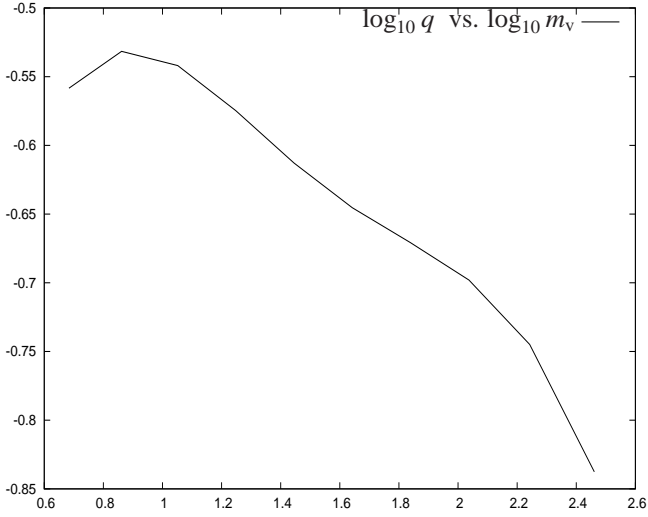}
\end{turn}
\caption{The logarithm$ {}_{10} $ of the primordial phase-space density
 $ q = (Z \; Q_{halo})/({\rm keV})^4 $
vs. the common logarithm of the virial mass of the galaxy 
$ m_{\rm v} \equiv M_{virial}/[10^{11} M_\odot]. \; q $ is obtained by
solving  eq.(\ref{eqq}).}
\label{x}
\end{figure}

The DM profile eq.(\ref{perf}) decreases with the characteristic 
length $ r_{lin} $ which is of the same order of magnitude than the halo radius $ r_0 $
in the empiric density profile eq.(\ref{bur}). We define 
the coefficient $ \alpha $ as $ \alpha \equiv r_{lin}/ r_0 $ and
determine it by fitting the linear profile to the Burkert profile in 
 \ref{perflin}.
The value of $ \alpha $ turns to be between 0.4 and 0.8 depending on the DM
particle statistics (see Table B.1).

Using the numerical values of the parameters eqs. (\ref{val1}) and 
(\ref{val2}), this theoretical formula takes the form
\be\label{muteo2}
\mu_{0 \, lin} = 391.1 \; \; \frac{({\rm MeV})^3}{\hbar^2 \; c^4} 
\; \frac{b_1}{\alpha} \; q_p^\frac{n_s}6
\; \int_0^{\infty} \gamma \; N(\gamma) \; d\gamma \; ,
\ee
where 
\be\label{defq}
q_p \equiv \frac{Q_p}{({\rm keV})^4} \; \hbar^3 \; c^8 \quad , 
\ee
and
\be\label{Nq}
N(\gamma)=\gamma^{n_s/2-1} \; 
\ln\left(d_0 \; q_p^\frac13 \; \gamma \right) \; T(\gamma) 
\quad , \quad  d_0 = 556.7 \; . 
\ee
From now on we use the dimensionless primordial density $ q_p $.

We identify the observed surface density  $ \mu_{0 \, obs} $ with 
the theoretical value obtained in the linear approximation 
$ \mu_{0 \, lin} $. 
We thus obtain the following trascendental equation in the variable $ q_p$:
\be\label{eqx}
q_p^\frac{n_s}6 \; \int_0^{\infty} \gamma \; 
N(\gamma) \; d\gamma =  \frac{\alpha}{b_1} \;
\frac{\mu_{0 \, obs} \; \hbar^2 \; c^4}{391.1 \; \; ({\rm MeV})^3} \; .
\ee
We compute the quantities in eq.(\ref{eqx}) using $ N(\gamma) $ eq.(\ref{defq})
[i. e. the transfer function $ T(\gamma) $] from the solution of the linearized 
Boltzmann-Vlasov equation obtained in \citep{ds1,gil}, so that:
\bea\label{intTg}
&& \int_0^{\infty} \gamma^{n_s/2} \; T(\gamma) \; \ln \gamma \; d\gamma =1.315\ldots
\; , \cr \cr
&&\int_0^{\infty} \gamma^{n_s/2} \; T(\gamma) \; d\gamma = 2.666\ldots
\eea
and hence,
\be\label{intNc}
\int_0^{\infty} \gamma \; N(\gamma) \; d\gamma = 
18.17 \left(1 + 0.0489 \; \ln q_p\right) \; .
\ee
These values correspond to fermions decoupling ultrarelativistically
at thermal equilibrium. Bosons and particles obeying the Maxwell-Boltzmann statistics
yield similar results as one sees from figs. \ref{tk} and \ref{3perf}.

For fermions decoupling ultrarelativistically at thermal equilibrium,
eq.(\ref{eqx}) takes then the form:
\be\label{eqq}
q_p^{0.161} \; \left(1+0.04891 \; \ln q_p\right) = \frac1{b_1} \; 
\frac{\mu_{0 \, obs} \; \hbar^2 \; c^4}{10330 \; ({\rm MeV})^3} \; .
\ee
where we used the numerical values in eqs.(\ref{val2}) and (\ref{intNc}).
The value of $ b_1 \sim 1 $ which provides the best fit to
the halo radius is $ b_1 \simeq 0.8 $ (see appendix \ref{perflin}).

We proceed now to solve numerically eq.(\ref{eqq}) 
to obtain the primordial
phase-space density $ q_p$ for the different values of $ \mu_{0 \, obs} $
given in Table 1.

\section{The DM particle mass and the decoupling temperature
from the galaxy surface density}

We plot in fig. \ref{x} the solution of eq.(\ref{eqq}), 
$ \; q_p\; $ vs. $ m_{\rm v} $. From eqs.(\ref{defZ}) and (\ref{defq})
$ q_p$ can be expressed as
\be\label{qfz}
q_p= \frac{Z \; Q_{halo}}{({\rm keV})^4} \; \hbar^3 \; c^8 \; .
\ee
Therefore, for a galaxy of mass $ m_{\rm v} $ the observed values of the phase-space density
$ Q_{halo} $ (fig. \ref{Q}) yields the factor 
$ Z $ as a function of the virial mass $ m_{\rm v} $ [eq. (\ref{mvir})].

\begin{figure}
\begin{turn}{-90}
\psfrag{"QG11.dat"}{$ \log_{10} Q^{-1}_{halo} $ vs. $ m_{\rm v} $}
\psfrag{"Z11.dat"}{$ \log_{10} Z $ vs. $ m_{\rm v} $}
\includegraphics[height=9.cm,width=9.cm]{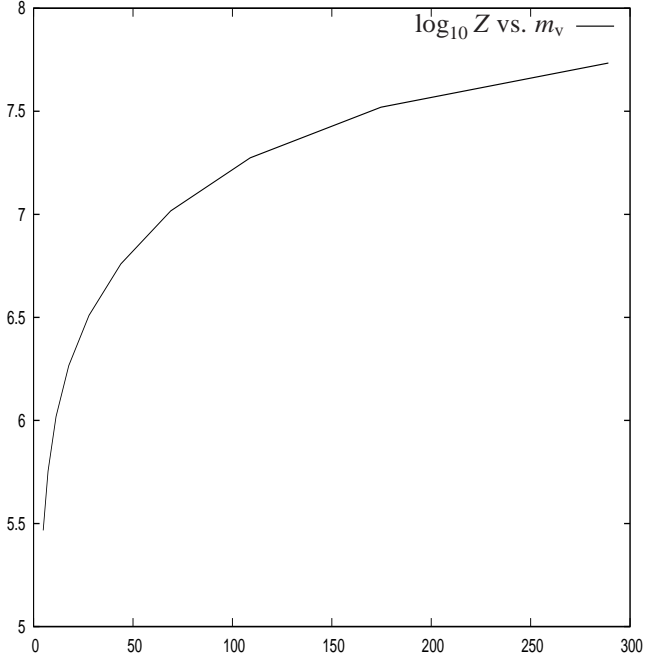}
\end{turn}
\caption{The common logarithm of the self-gravity decreasing 
factor $ Z $ computed from eq.(\ref{qfz}) with $ q_p $ solution 
of eq.(\ref{eqq}) [fig. \ref{x}]  vs. the virial 
mass of the galaxy $ m_{\rm v} \equiv M_{virial}/[10^{11} M_\odot] $.}
\label{QZ}
\end{figure}

In Fig. \ref{QZ} we plot $ \log_{10} Z $ vs. $ m_{\rm v} $, and $ \log_{10} 
Q_{halo}^{-1} $ vs. $ m_{\rm v} $ is plotted in Fig. \ref{Q}.
We see that $ Q_{halo} $ decreases with $ m_{\rm v} $ while 
$ Z $ increases with $ m_{\rm v} $
in such a way that the product $ Z \; Q_{halo} $ 
is roughly {\bf constant}. Moreover, as follows from 
eqs.(\ref{defZ}) and (\ref{002}) $ Z \; Q_{halo} $ gives the DM
particle mass 
\be\label{m4fd}
 m^4 = 49.0 \; Z \;  Q_{halo} \; .
\ee
We notice in fig. \ref{QZ} that the factor $ Z $ changes by about two orders of magnitude 
$$ 
2.9 \; 10^5 \lesssim Z \lesssim 5.4 \; 10^7  \; ,
$$
over a large range of values of the virial mass. The variation of $ Z $ is relevant
in the context of galaxy formation but not for the particle DM determination.
Since $ m $ goes as $ Z^{1/4} $ even a large change in $ Z $ merely produces a
small change in $ m $. For example, changing $ Z $ by a factor 100 changes $ m $ by a factor
$ 3.2 $. 

\medskip

\begin{figure}
\begin{turn}{-90}
\psfrag{"eme11.dat"}{$ m $ in keV vs. $ m_{\rm v} $}
\includegraphics[height=9.cm,width=9.cm]{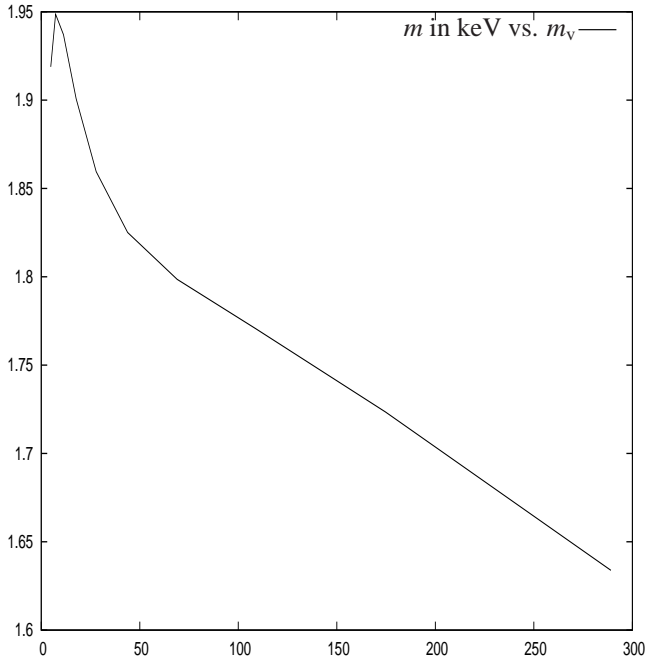}
\end{turn}
\caption{The DM particle mass $ m $ in keV following eq.(\ref{dm})
with the values of $ q_p$ solution of eq.(\ref{eqq})
vs. the virial mass of the galaxy 
$ m_{\rm v} \equiv M_{virial}/[10^{11} M_\odot] $.
We see that the DM mass $ m $ exhibits the same variation with $ m_{\rm v} $ 
than the surface density $ \mu_{0 \, obs} $ in fig. \ref{muo}. 
The precision in the observations 
of the surface density $ \mu_0 $ translates on the precision
of the DM mass $ m $. A value for $ m $ slightly below 2 keV is favoured.}
\label{eme}
\end{figure}

We obtain the DM particle mass $ m $ from eqs.(\ref{Qprim})-(\ref{002}) 
in terms of the invariant phase-space density $ Q_p $:
\be\label{dm}
m = m_0 \; \frac{Q_p^\frac14}{{\rm keV}} = m_0 \; q_p^\frac14
\quad , \quad 
m_0 \equiv \left(\frac2{g}\right)^\frac14 \; 
\frac{\sqrt{\pi}}{I_2^\frac58} \; 
\left(\frac{I_4}3\right)^\frac38 \; {\rm keV} \; ,
\ee
where
\bea\label{mcero}
&& m_0 = 2.6462 \; {\rm keV}/c^2 \quad 
{\rm for ~ Dirac ~fermions} \quad , \cr \cr
&& m_0 = 2.6934 \; {\rm keV}/c^2  \quad 
{\rm for ~scalar ~ Bosons.} 
\eea
The numerical coefficients here correspond to ultrarelativistic 
decoupling 
at thermal equilibrium. For decoupling out of thermal equilibrium the 
coefficients are of the same order of magnitude \citep{dm2}.

In fig. \ref{eme} we plot $ m $ according to  eq.(\ref{dm}) 
with the values 
of $ q_p$ solution of eq.(\ref{eqq}) (fig. \ref{x})
and $ \mu_{0 \, obs} $ given in Table 1.
The precision in the observations 
of the surface density $ \mu_0 $ translates on the precision
of the DM particle mass $ m $.

\bigskip

We find $ m $ about 2 keV (up to $ \pm 10 $\%) for $ b_1 = 0.8 $.
More generally, $ m $ is in the keV {\bf scale} for $ b_1 \sim 1 $.

\bigskip

The variation of the observed surface density $ \mu_{0 \, obs} $ 
with the core radius $ r_0 $ (fig. \ref{muo}) is similar to: 
\begin{itemize}
\item{(a) the variation of the DM particle mass $ m $ displayed
in fig. \ref{eme},}
 \item{(b) the variation of the primordial phase-space density
$ q_p$ in fig. \ref{x},}
\item{(c) the variation of
the  density contrast in fig. \ref{con}. }
\end{itemize}
Therefore, the precision in the observations of the surface density 
$ \mu_0 $
translates on the precision in the evaluation of the DM mass $ m $.

\bigskip

From the solution for $ q_p$ eq.(\ref{eqq}) and 
fig. \ref{x} we can also compute the 
number of ultrarelativistic degrees of freedom at decoupling 
$ g_d $ and therefore
the decoupling temperature $ T_d $ which is a further relevant 
characteristic magnitude of the DM particle.
For Dirac fermions decoupling ultrarelativistically 
at thermal equilibrium
the number of ultrarelativistic degrees of freedom at decoupling
can be expressed from eq.(\ref{gdu}) as 
\be\label{gdu2}
g_d = 1365.5 \; q_p^\frac14  \; .
\ee
And from fig. \ref{x}:
\be\label{rangoq}
0.14 < q_p< 0.3 \quad , \quad 0.61 < q_p^\frac14 < 0.74 \; .
\ee
We thus find that for thermal fermions $ g_d $ is in the interval
$$
833 <  g_d < 1010 \qquad {\rm thermal ~ fermions} \; , 
$$
which correspond to physical decoupling temperatures 
[eq.(\ref{temp})] above 100 GeV.

The gravitino is a popular DM candidate decoupling at thermal
equilibrium which can provide such values of $ g_d \sim 1000 $ 
in non-minimal supergravity extensions of the standard model of particle physics.
(In the minimal supersymmetric extension of the
standard model (MSSM) one has the value $ g_d = 228.75 $ \cite{gkr,stef}).

For DM particles decoupling out of thermal equilibrium as sterile
neutrinos, the primordial power spectrum and therefore the inferred
values for the mass of the DM particle change by a factor of order one
\cite{dm1,boy,dm2,dm3,pet,viel}. The low-momentum regime is enhanced in 
the out of equilibrium particle distributions $ F_d(y) $ \cite{dm1}
and therefore the dimensionless momentum  $ I_2 $ of  $ F_d(y) $
is smaller for out of equilibrium decoupling than for thermal 
equilibrium decoupling.
As a consequence, we see from eq.(\ref{masa}) that we can have smaller $ g_d $ 
for smaller $ I_2 $ always keeping $ m $ in the keV scale.

Sterile neutrinos which decouple out of equilibrium are today the
front-runner candidate for WDM in the keV mass scale.

In summary, the DM particle mass is in the keV scale whether the DM
particle decouples in or out of thermal equilibrium. The fact that the 
DM particle mass is in the keV scale is a robust result
which does not depend on the details of the particle physics models.
Of course, to fix the number  within  the scale  $ 1 < m < 10 $ keV depends on
the details of the particle model. Our aim in this paper
is not to analyze the observational constraints on
the DM particle models but to determine the DM particle mass scale from 
general fundamental grounds and observations.

\section{Non-universal structural galaxy properties}

We compute here for illustration non-universal galaxy quantities as the halo radius, 
galaxy mass, halo central density and squared halo velocity. 
These calculations are independent of determination of the DM particle mass
and are presented to see what kind of results provide the linear approximation.
Let us anticipate that the linear approximation for non-universal galaxy properties
agrees with the observed values within one order of magnitude.

Notice that
our determination of the DM particle mass does not relay to
these non-universal galaxy quantities.

\medskip

\begin{figure}
\begin{turn}{-90}
\psfrag{"r0lin11.dat"}{$ r_0 $ vs. $ m_{\rm v} $}
\psfrag{"r0mvir11.dat"}{$ r_{0} $ data vs. $ m_{\rm v} $}
\includegraphics[height=9.cm,width=9.cm]{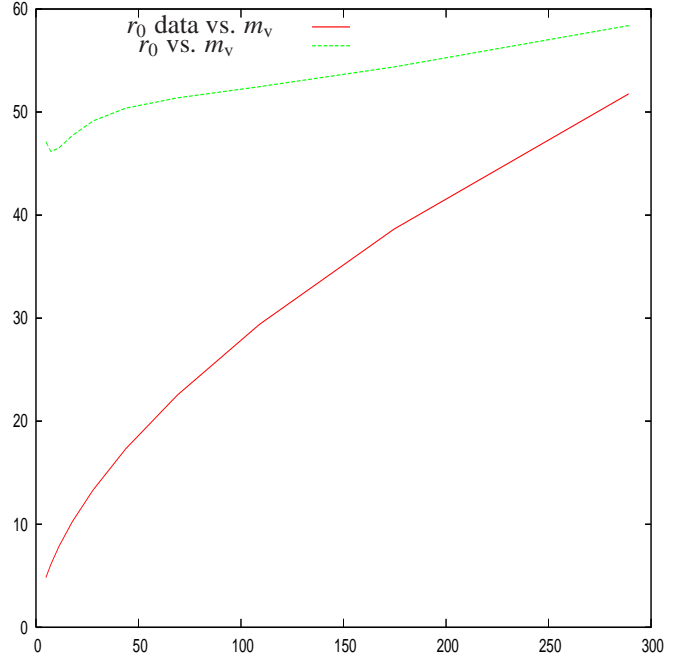}
\end{turn}
\caption{The computed halo radius 
$ r_0 $ in kpc from eq.(\ref{rlin3})in broken green line,
the halo radius $ r_0 $ in kpc from the real data in Table 1 in solid red line
vs. the virial mass of the galaxy $ m_{\rm v} \equiv M_{virial}/[10^{11} M_\odot] $.
The theoretical $ r_0 $ computed from first principles approaches asymptotically
the observed $ r_0 $ for large galaxies.}
\label{rlin}
\end{figure}

The characteristic length of the linear profile 
$ r_{lin} $ eq.(\ref{rlin1})
takes the following form in terms of $ q_p$ eq.(\ref{defq}):
\be\label{rlin2}
r_{lin} = 21.1 \; q_p^{-\frac13} \; \; {\rm kpc} \; .
\ee
In fig. \ref{rlin} we plot
$ \; r_{lin} $ from eq.(\ref{rlin2}) and $ \alpha \; r_0 $
from the data in Table 1 as functions of $ m_{\rm v} $. 

\medskip

The halo radius in the linear approximation is given by 
$ r_{lin} = \alpha \; r_0  $ which for DM Dirac fermions becomes
\be \label{rlin3}
r_0 \equiv \frac{r_{lin}}{0.688} = 
30.7 \; q_p^{-\frac13} \; \; {\rm kpc} \; ,
\ee
where we used $ \alpha = 0.688 $  obtained in appendix \ref{perflin}
by fitting the Burkert and linear profiles.

Using the range of values of $ q_p$ eq.(\ref{rangoq}) 
obtained by solving eq.(\ref{eqq}) yields
$$
46 \; {\rm kpc} \; < r_0 < 59  \; {\rm kpc} \; .
$$
which is in the upper range of the observed $ r_0 $ values in Table 1.
Namely, the linear approximation for the halo radius give 
values above or in the range of the observations.

\medskip

\begin{figure}
\begin{turn}{-90}
\psfrag{"mgal11.dat"}{$ \log_{10} [M_{gal}/M_{virial}] $ vs. $ m_{\rm v} $}
\includegraphics[height=9.cm,width=9.cm]{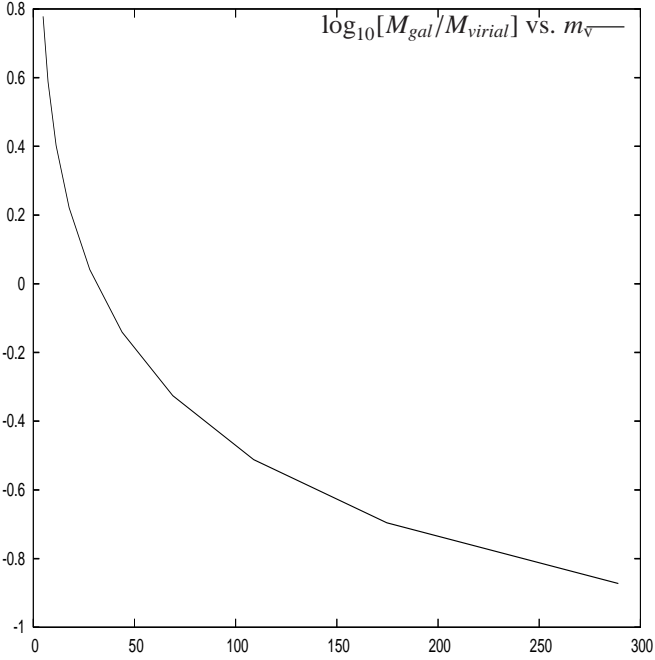}
\end{turn}
\caption{The common logarithm of the predicted total mass of the galaxy 
$ M_{gal} $ given by eq.(\ref{mgal}), divided by the 
observed virial mass $ M_{virial} $ vs. the observed virial mass of 
the galaxy $ m_{\rm v} \equiv M_{virial}/[10^{11} M_\odot] $.
The ratio $ M_{gal} /  M_{virial} $ turns to be in the interval
$ 0.12 < M_{gal}/M_{virial} < 5.0 $. 
Notice that the difference of $ M_{gal} $ with
$  M_{virial} $ is irrelevant to the determination of the DM particle mass.} 
\label{mgmv}
\end{figure}

The total mass of the galaxy $ M_{gal} $ follows by integrating the 
density profile eq.(\ref{perf}). We find 
\be\label{mgal}
 M_{gal} \simeq 20 \; r_0^3 \; \rho_{lin}(0) 
= 20 \; r_0^2  \; \mu_{0 \, obs} \; .
\ee
In fig. \ref{mgmv} we plot $ \; M_{gal}/M_{virial} $ vs. $ m_{\rm v} $ 
where the observed $ m_{\rm v} $ and $ M_{virial} $ are defined by eq.(\ref{mvir}).

We see that the ratio $ M_{gal} /  M_{virial} $ turns to be in the interval,
$$
0.12   < \frac{M_{gal}}{M_{virial}}< 5. \; .
$$

\begin{figure}
\begin{turn}{-90}
\psfrag{"con11.dat"}{$ \rho_{lin}(0)/{\bar\rho}_{DM} $ vs. $ m_{\rm v} $}
\includegraphics[height=9.cm,width=9.cm]{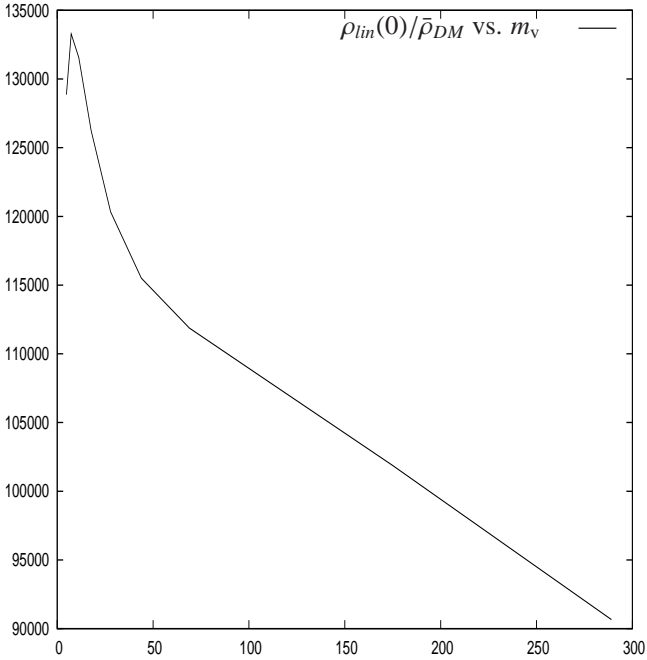}
\end{turn}
\caption{The ratio  $ \rho_{lin}(0)/ {\bar\rho}_{DM} $
between the maximum DM mass density $ \rho_{lin}(0) $
and the average mass density in the universe $ {\bar\rho}_{DM} $ 
vs. the virial mass of the galaxy $ m_{\rm v} \equiv M_{virial}/[10^{11} M_\odot] $.
The ratio $ \rho_{lin}(0)/ {\bar\rho}_{DM} $ turns to be between 1/3 and 1/2 
of the observed value $ \sim 3 \times 10^5 $ \citep{asp}.}
\label{con}
\end{figure}

\medskip

The contrast
density, that is, the ratio between the maximum DM mass density $ \rho_{lin}(0) $
and the average DM mass density $ {\bar\rho}_{DM} $ in the universe results
$$
{\rm contrast} \equiv \frac{\rho_{lin}(0)}{{\bar\rho}_{DM}}
$$
with $ {\bar\rho}_{DM} = \Omega_{DM} \;  \rho_c $ and $ \Omega_{DM} $ and $ \rho_c $
given by eq.(\ref{val0}). $ \rho_{lin}(0) $ is given by eq.(\ref{dfmulin}) as
$$
\rho_{lin}(0) = \frac{\mu_{0 \, lin}}{r_0}   \; .
$$
We plot in fig. \ref{con} the contrast density
\be\label{contra}
{\rm contrast} = \frac{\mu_{0 \, lin}}{\Omega_{DM} \;  \rho_c \; r_0} 
\ee
As seen from fig. \ref{con}, the ratio obtained is between 1/3 and 1/2 of the observed 
value $ \sim 3 \times 10^5 $ in \citep{asp}. The values obtained are 
below the observed values
because the linear halo radius $ r_0 = r_{lin}/0.688 $ is larger
than the observed halo radius $ r_0 $ and the density contrast goes as 
$ 1/ r_0 $ eq.(\ref{contra}). 
This property shows again that the larger and more dilute
is the galaxy the better is the linear approximation 
for non-universal quantities (see Table 2). 

\medskip

Notice that we consider the whole range of galaxy virial masses
going from 5 to 300 $ \times 10^{11} \; M_\odot $.
Universal quantities as the surface density stay constant up to 
$ \pm 20 \% $ within this wide range of galaxy masses.

\medskip

It is relevant to evaluate the halo velocity given by eq.(\ref{vhprom}) 
\be\label{v2halo}
 {\overline {v^2}}_{halo} = 2.316 \; G \; \mu_0 \; r_0 \; .
\ee
Using eq.(\ref{rlin3}) this equation becomes
\be\label{vhalo}
{\overline {v^2}}_{halo} = 6.705 \; \frac{\mu_0}{{\rm MeV}^3} \;  
({\rm km/sec})^2  \; q_p^{-\frac13} \; .
\ee
From Table 1 the input observed surface density takes the value 
\be\label{mu0ob}
\mu_0 \simeq 6000 \; {\rm MeV}^3 \; .
\ee 
Eq.(\ref{vhalo}) thus becomes
\be\label{vq}
\sqrt{{\overline {v^2}}}_{halo \; lin} = 
\frac{201}{q_p^{\frac16}} \; {\rm km/sec} \; .
\ee
The obtained range of values of $ q_p$ eq.(\ref{rangoq}) yields
$ q_p^{\frac16} \simeq 0.77 $ and 
\be\label{vkev}
\sqrt{{\overline {v^2}}}_{halo \; lin} \simeq 260 \; {\rm km/sec} \; .
\ee
This value is to be compared with the values arising 
from $ \mu_0 $ and eq.(\ref{v2halo}) and the observed values 
$ r_0 $ in Table 1.
\be\label{vobs}
79.3  \; {\rm km/sec} < \sqrt{{\overline {v^2}}}_{halo} < 
261 \; {\rm km/sec} \; .
\ee
The halo central density in the linear approximation is given from 
eqs. (\ref{rlin3}) and (\ref{mu0ob}) by
$$
\rho_{0 \, lin} = \frac{\mu_0}{r_0} = 2.90 \; 10^{-25}  \; q_p^{\frac13} 
\; \frac{\rm g}{{\rm cm}^3}  \; .
$$
Using the range of values of $ q_p $ eq.(\ref{rangoq}) obtained by solving eq.(\ref{eqq}) 
yields 
$$
1.33  \; 10^{-25}  \; \frac{\rm g}{{\rm cm}^3} < \rho_0 < 
1.94 \; 10^{-25}  \; \frac{\rm g}{{\rm cm}^3} \; ,
$$
for $ 1.6 \; {\rm keV} < m < 1.9 $ keV, which must be compared with the observed values  
of $ \rho_0 $ given in Table 2.

We see that the linear approximation produces halo central densities smaller or
in the range of the observations and halo velocities larger than the observed ones
by a factor of order one. 

\medskip

 Clusters of galaxies exhibit halo radius $ r_0 $ about 210 kpc \cite{bs} well 
beyond the linear halo radius $ \sim 50$ kpc. Hence, clusters of galaxies cannot 
be described by the initial conditions used here. 
Chosing general random 
fields  $ g(\vk) \neq 1 $ fulfilling eq.(\ref{gg}) will provide general configurations
with a large range of masses and sizes. Each realization of the random field $ g(\vk) $ 
produces a possible galaxy configuration. The factor $ g(\vk) $ multiplies the transfer
function $ T(k) $ and therefore is to be added in the r. h. s.
of eqs.(\ref{flueq}), (\ref{defN}) and (\ref{Nq}) and inside the $\vk-$integrands [r. h. s. of 
eqs.(\ref{perf}), (\ref{perfun}), (\ref{muteo}), (\ref{muteo2}), 
(\ref{eqx}), (\ref{intTg}) and (\ref{intNc})].

\begin{figure}
\includegraphics[height=9.cm,width=9.cm]{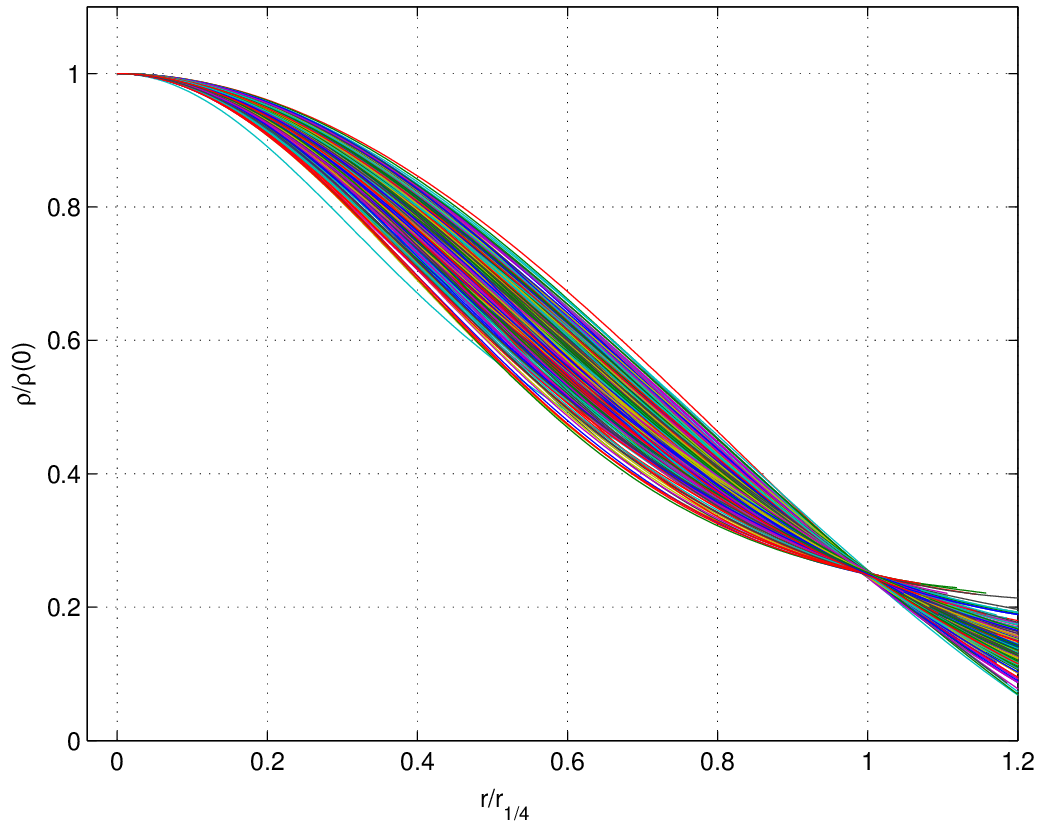}
\caption{The normalized density profile $ \rho_{lin}(r)/\rho_{lin}(0) \; $
averaged in the angles for 500 random initial conditions $ g({\vec k}) $
vs. $ r/r_{1/4} $ \cite{CD}. $ r_{1/4} $ being the point where  $ \rho_{lin}(r) $ takes
1/4 of its value at the origin.  $ r_{1/4} $ coincides with the halo radius 
in the Burkert profile.}
\label{perfCD}
\end{figure}

We plot the density profiles in figs. \ref{plgg} and \ref{perfCD}.
Fig. \ref{perfCD} displays 500 profiles averaged in the angles for random initial conditions.
One can see that the random initial fluctuations only produce mild changes in the shape 
of the density profiles. Therefore, restricting ourselves for simplicity
to initial primordial conditions with $ g(\vk) \equiv 1 $ still provides relevant
physical results. 

\begin{table*}
 \centering
 \begin{minipage}{140mm}
\begin{tabular}{cccc} \hline  
      & Observed Values & Linear Theory & Wimps in linear theory \\
\hline 
  $ r_0 $ & $ 5 $ to $ 52 $ kpc &  $ 46 $ to $ 59 $ kpc & $ 0.0045 $ pc \\
\hline 
  $ \rho_0 $ & $ 1.57  $ to $ 19.3 \times 10^{-25}  \; \frac{\rm g}{{\rm cm}^3} $  & 
$ 1.33  $ to $ 1.94  \times 10^{-25}  \; \frac{\rm g}{{\rm cm}^3} $  &
$ 1.773  \times  10^{-14}  \; \frac{\rm g}{{\rm cm}^3} $ \\ \hline 
  $ {\sqrt{{\overline {v^2}}}_{halo}} $ &  79.3  to  261   km/sec & 
 260   km/sec &  0.0768   km/sec \\
\hline   
\end{tabular}
\caption{Non-universal galaxy quantities from the observations (Table 1
combined with the virial) and from the linear theory results.
The corresponding dark matter particle mass is plotted in fig. \ref{eme}
and is in the range $ 1.6 - 1.9 $ keV.
The larger and less denser are the galaxies, 
the better are the results from the linear
theory for non-universal quantities. The last column corresponds 
to 100 GeV mass wimps.
The wimps values strongly disagree by several orders of 
magnitude with the observations.}
\end{minipage}
\end{table*}

\section{The density profile: cores vs. cusps}

The properties of the density profile $ \rho_{lin}(r) $ depend
on the free streaming length $ r_{lin} $ 
and therefore on the mass of the DM particle as we discuss here below.

We find from eqs.(\ref{dfmulin}), (\ref{muteo2}) and (\ref{intNc})
for the density profile at the origin
\be\label{coc1}
\hskip -1cm \rho_{lin}(0) = \frac{\mu_0}{r_0} = 336.7 \; b_1 \; 
q_p^{\frac{n_s+2}6} \times 
\left[ 1 + 0.04891 \; \ln q_p\right] \;
\frac{({\rm MeV})^3}{{\rm kpc}} \; .
\ee
We use from eqs.(\ref{dm}) and (\ref{rlin3}) that
\be\label{qrlin}
q_p= \left(\frac{m}{m_0}\right)^4 \quad , \quad
 r_{lin} = 77.23 \; {\rm kpc} \; \left(\frac{\rm keV}{m}\right)^\frac43 \; ,
\ee
for DM particles decoupling ultrarelativistically at thermal equilibrium
with $ m_0 $ given by eq.(\ref{mcero}). Then eq.(\ref{coc1}) can be written as 
\bea\label{coc2}
&&\rho_{lin}(0) = 1.622 \; 10^{-25} \; \left(\frac{m}{1.75 \; {\rm keV}}\right)^{1.976}
\cr \cr 
&& \times \left[ 1 + 0.2428 \; \ln \left(\frac{m}{1.75 \; {\rm keV}}\right)\right] \; 
\frac{\rm g}{{\rm cm}^3} \; ,
\eea
where we used the numerical values from eqs. (\ref{val2}) and (\ref{mcero})
and the conversion of units:
$$
\frac{({\rm MeV})^3}{{\rm kpc}} = 0.1483698 \; 10^{-26} \; \frac{\rm g}{{\rm cm}^3} \; .
$$

For the DM particle mass value $ m \sim 2 $ keV found in the previous
section, $ \rho_{lin}(0) $ from eq.(\ref{coc2}) is two to three times
smaller than the observed values (as it is the contrast density, 
discussed in the previous section). This is
not surprising because $ \rho_{lin}(0) $ is not an universal quantity
and given the approximation of our theoretical computation.

We derive in \ref{asypro}, eq.(\ref{asidur}) the density profile behaviour 
for $ r \gtrsim r_{lin} $ where $ r_{lin} $ is given by eq.(\ref{qrlin}):
\bea\label{asidur2}
&&\rho_{lin}(r\gtrsim r_{lin}) =
10^{-26} \;  \frac{\rm g}{{\rm cm}^3} \;
\left(\frac{36.45 \; {\rm kpc}}{r}\right)^{1.482} \; \cr \cr
&&\times \ln\left(\frac{7.932\; {\rm Mpc}}{r}\right) \;
\left[ 1 + 0.2417 \; \ln \left(\frac{m}{\rm keV}\right)\right] \; .
\eea
It should be noticed
that this behaviour has only a mild logarithmmic dependence on the DM particle
mass $ m $. The scales in eq.(\ref{asidur2}) only depend on known cosmological
parameters and not on $ m $.

\begin{figure}
\begin{turn}{-90}
\psfrag{"rhoBup.dat"}{Burkert profile}
\psfrag{"rhol1LG.dat"}{$ \rho_{lin}(r) $ for m = 1 keV}
\psfrag{"rhol2LG.dat"}{$ \rho_{lin}(r) $ for m = 2 keV}
\includegraphics[height=9.cm,width=9.cm]{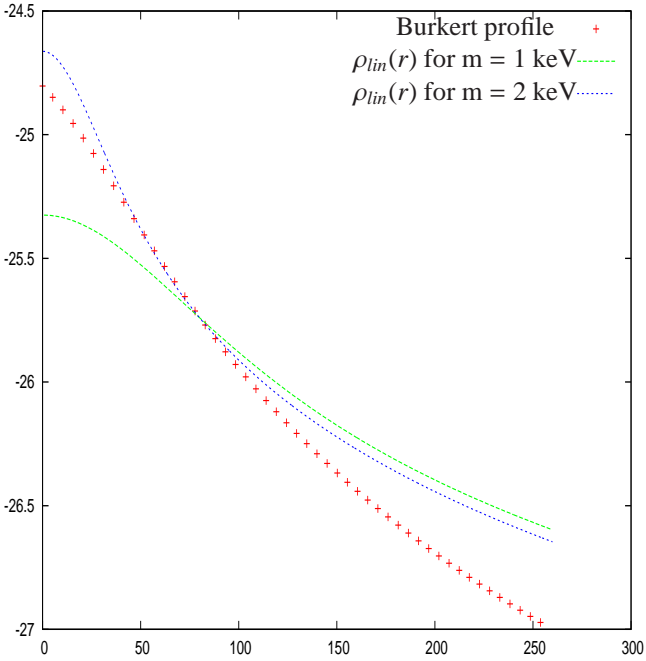}
\end{turn}
\caption{The common logarithm of the density profile 
$ \rho_{lin}(r) $ according to eqs.(\ref{perf}) and (\ref{perfun}) 
in g/${\rm cm}^3 $ vs. $ r $ in kpc and the Burkert profile
eq.(\ref{bur}). The Burkert profile is plotted with red crosses 
for the largest galaxy 
in Table 1 with $ r_0 = 51.8 $ kpc and $ \rho(0) = 1.57
\times 10^{-25}  \; \frac{\rm g}{{\rm cm}^3} $. Notice that the 
agreement of the linear density profile $ \rho_{lin}(r) $ 
with the Burkert profile is best for a DM particle mass slightly below 2 keV.}
\label{plgg}
\end{figure}

We plot in fig. \ref{plgg} the density
profile $ \rho_{lin}(r) $ according to eqs.(\ref{perf}) and (\ref{perfun}) 
for DM particle masses $ m $ of 1 and 2 keV and the Burkert
density profile for the largest galaxy $ r_0 = 51.8 $ kpc and $ \rho(0) = 1.57
\times 10^{-25}  \; \frac{\rm g}{{\rm cm}^3} $ in Table 1.
We see from  fig. \ref{plgg} that the density profile $ \rho_{lin}(r) $
best follows the Burkert profile for a DM particle
mass $ m $ slightly below 2 keV. This is in agreement with Fig. \ref{eme}
for the DM particle mass where a value for $ m $ slightly below 2 keV is favoured.

\bigskip

We present in this paper clear evidences for a DM particle
mass in the keV scale. However, one can wonder what is the shape 
of the density profile and the value of the density at the origin
for a typical hundred GeV wimp. 

Since wimps are supposed to decouple non-relativistically, 
eq.(\ref{Qprim}) does not apply to them. For DM particles 
decoupling non-relativistically $ Q_p $ is given by \citep{dm2}
\be\label{qpnr}
Q_p = \frac{\Omega_{DM} \; \rho_c}{2 \; T_{\gamma}^3} \; g_d \;
(m \; T_d)^\frac32 \quad {\rm nonrelativistic \; decoupling} \; .
\ee
For a $ 100 $ GeV wimp decoupling at the typical
temperature $ T_{d \, wimp} = 5 $ GeV, we find from 
eqs.(\ref{defq}) and (\ref{qpnr})
\be\label{qwimp}
q_{p \; wimp} = 0.3166 \; 10^{21} 
\ee
where we used that $ g_d \simeq 80 $ at such decoupling temperature 
\citep{kt}. We then find from eq.(\ref{coc1}) the central 
density value $ \rho_{lin}(0) $ for such value of $ q_p$:
\be\label{rho0w}
\rho_{lin}(0)_{wimp}  \simeq 1.773 \times 10^{-14} \; 
\frac{\rm g}{{\rm cm}^3} 
\ee
This value for the wimps density profile at the origin turns to be 
larger than the observed values by {\bf eleven orders of magnitude}. 
This result indicates that the DM particle mass is not in the GeV scale.
DM particles at the keV scale reproduce very well 
both the surface density and the density profile at the origin.

\medskip

The free-streaming length $ r_{lin} $ is the characteristic scale where  
$ \rho_{lin}(r) $ varies (see fig. \ref{3perf}). 
This length is of the order of hundred kpc for keV mass scale
DM particles as shown by eq.(\ref{qrlin}). For a hundred GeV wimp 
decoupling at $ T_{d \, wimp} = 5 $ GeV we find from eqs.(\ref{rlin2}) 
and (\ref{qwimp})
\be\label{rlinW}
r_{lin}(m_{wimp}=100 \; {\rm GeV}, \; T_{d \, wimp} 
= 5\; {\rm GeV}) = 0.0031 \;  {\rm pc} = 639 \; {\rm AU} \; .
\ee
Therefore, with such small $ r_{lin} $ for wimps we can use for 
all relevant galactic scales the asymptotic behaviour of 
$ \rho_{lin}(r) $ eq.(\ref{asifin}) valid for $ r \gg r_{lin} $. That is,
\bea\label{rhow}
&&\rho_{lin}(r\gtrsim 0.003 {\rm pc})_{wimp}
= 0.8064 \; 
10^{-14} \;  \frac{\rm g}{{\rm cm}^3} \\ \cr
&&\times \left(\frac{0.0031 \; {\rm pc}}{r}\right)^{1.482}  \;
\left[ 1 + 0.04616 \; \ln \left(\frac{0.0031 \; 
{\rm pc}}{r}\right) \right] \; . \nonumber
\eea
This profile clearly exhibits a {\bf cusp} behaviour for scales 
$ 1 {\rm pc} \gtrsim r \gtrsim 0.003 $ pc.
Notice that this asymptotic formula eq.(\ref{rhow}) approximatively 
matches around $ r \sim 0.003 $ pc the value of the wimp profile at 
the origin eq.(\ref{rho0w}).

In summary, the density profile $ \rho_{lin}(r) $ eq.(\ref{perf}) 
exhibits a cusp around the origin 
for a wimp DM particle and a core behaviour at $ r = 0 $ for a keV 
scale DM particle mass.

\medskip

We display in fig. \ref{perT} the density profile
for 100 GeV wimps and the NFW profile for the largest galaxy in Table I.
The density profile for 1-2 keV particles in fig. \ref{plgg}
and the density profile for wimps in fig. \ref{perT}
practically coincide for $ r \gtrsim 30 $ kpc while they strongly 
differ at smaller scales ($ r \lesssim 30 $ kpc). 
The keV mass profile exhibits a core like
the Burkert profile while the wimp profile exhibits 
a cusp like the NFW profile.

In this way, the value of the mass of the dark matter particle 
turns to be between 1 and 2 keV, and the number of 
ultrarelativistic degrees of freedom 
of the dark matter coupling at decoupling $ g_d $, or similarly, the 
decoupling temperature $ T_d $ turns to be above 100 GeV.

\medskip

We can also evaluate the halo velocity for wimps from the general
formula eq.(\ref{vq}) and the value of $ q_{p \; wimp} $ 
 eq.(\ref{qwimp}). We obtain
$$
\sqrt{{\overline {v^2}}}_{halo \; lin \; wimp} = 0.0768 \; {\rm km/sec} 
$$
three orders of magnitude below the observed halo velocities 
eq.(\ref{vobs}).
Recall that keV scale DM particles yield a halo velocity 
eq.(\ref{vkev}) of the same order
of magnitude than  the observed halo velocities.
Therefore, keV DM particles may solve the problem in
the halo velocities recently noticed by \citep{LK} for the bullet cluster
when CDM wimps are used.

\medskip

The analytic expressions we derived for the density profile,
and the mass of the dark matter particle also imply that 
keV dark matter particles always produce cored density profiles while 
heavy dark matter particles as wimps 
($m = 100$ GeV, $ T_d $ = 5 GeV) inevitably produce cusped
profiles at scales of 0.003 pc. These results are independent
of the particle model and vary very little with the statistics of the dark matter 
particle.

\section{On the Validity of the Linear Approximation}\label{lineal}

The linear approximation to the Boltzmann-Vlasov equation
is valid as long as the density contrast is at most of order one.
However, in the non-linear regime the density fluctuations 
relevant to the galaxy profiles grow with time independently of the wavenumber.
Therefore, the shape of the linear profile survives in the non-linear regime.
Only the profile normalization changes according with the non-linear evolution.

These results from linear approximation provide in principle
only estimates since non-linear effects (including for instance mergers)
are expected to be important.
However, it turns out that the obtained linear results well reproduce the observations.

Of course, the theory of galaxy formation requires N-body simulations, beyond the
scope of this paper.

\medskip 

Notice that general arguments based on the Boltzmann-Vlasov equation
show that the cored or cusped character of a profile is preserved
through mixing and mergers and that
cusps do not become steeper neither shallower through  mixing and mergers
\citep{wd}.

Therefore, the cored and cusped character we find for 
the linear profiles depending on the DM particle mass considered 
(keV and GeV mass scale, respectively) should remain 
valid after mixing and mergers are taken into account.

\medskip

Moreover, recent $N$-body $\Lambda$CDM simulations (Acquarius)
have found that the DM halos form in a sort of "monolithical" way \cite{wang}.
Their inner regions,  that contain the visible galaxies, are found  
to be stable since early times and  contrary to previous believes, 
major mergers (i.e. those 
with progenitor mass ratios greater than 1:10)  are found to contribute 
little to their total mass growth \cite{wang}. This indicates that nonlinearities 
(i.e. mergers) have a reduced importance.
Minor mergers, secondary infall, rare major mergers are certainly important for 
details, but the essential features of DM halos are determined during the 
fast-accretion phase of their gravitational collapse, as  the 
history of the quasar-galaxies coevolution also seems to indicate \cite{dane}.

The halo formation essentially consists of two main phases: A first
fast accretion phase (that can be treated by the linear approximation),
and a second subsequent slow accretion phase with mergers and infalls, 
that have a random character and that can only be described by numerical
simulations. This second phase does not have an essential influence in the
shape of the halo profile. Thus, in order to explain the observed halo
profiles one just needs to describe the first phase of halo formation, as
we do here in this paper. 

\medskip

Evidence based on the phase space density pointing towards a DM particle mass
in the keV scale was presented in refs. \cite{dm1,dm2}.
Notice in this respect that the linear fluctuations as well as the spherical
model (which contains the nonlinearities) both give values for the DM particle mass 
in the keV scale which only differ by a factor ten. 

\medskip

Analytic methods have been used to derive galaxy properties
using the primordial power of the density fluctuations (see for example \cite{hs,pjep})
and using the spherical model  \citet{EBa,EBb}. 

\bigskip

In summary, the solution of the linearized Boltzmann-Vlasov equation 
presented here provides a satisfactory picture of the {\bf general} 
galaxy properties. Although nonlinear effects and baryons are not 
taken into account, the linear description presented here qualitatively 
reproduces the main non-universal and general
characteristics of a galaxy summarized in Table 2. 
Moreover, the agreement is even quantitative 
(approximatively) for the linear halo radius $ r_0  $, 
the galaxy mass $ M_{gal} $,
the linear  halo central density $ \rho_0 $ and the halo velocity 
$ {{\overline {v^2}}_{halo}}^{1/2}_{lin} $ compared to the 
respectived observed 
values in the limiting case of large galaxies (both $ r_0 $ 
and $ M_{gal} $ large). 
The agreement is very good for universal galaxy quantities 
as the surface density and the density profile as discussed above.

\medskip

The linear approximation for the density fluctuations amplitude 
today is clearly only an estimate for the true nonlinear value.
However, the DM particle mass derived from the phase-space density
in the linear approximation only differs by one order
of magnitude from the nonlinear value obtained from the spherical model
\cite{dm2}.

\medskip

Interestingly enough, it is possible to derive the value of the surface density $ \mu_0 $
from CDM simulations. Values of the product $ r_s \; \rho_s $ from NFW fits to CDM simulations
for galaxies were reported in \cite{yh}. From these values of $ r_s \; \rho_s $ we can derive
the surface density $ \mu_0 $, since $ \mu_0 = \rho_0 \; r_0 \simeq 25 \; r_s \; \rho_s $ 
with the result 
\be\label{mucdm}
\mu_0^{CDM} \simeq 10^7 \;  M_\odot /{\rm pc}^2 \; .
\ee
[Notice that $ \rho_s $ in \cite{yh} differs by a factor four from eq.(\ref{nfw})].

We see that the surface density from CDM simulations is {\bf five orders} 
of magnitude {\bf larger}
than the observed surface density $ \mu_{0 \, obs} \simeq 120 \; M_\odot /{\rm pc}^2 $
\citep{kor,dona,span}. 

It is illuminating to insert in eq.(\ref{eqq}) the above value of the CDM surface density 
$ \mu_0^{CDM} $ eq.(\ref{mucdm})instead of the observed value $ \mu_{0 \, obs} $. 
This gives for the mass of the CDM particle 
$ m^{CDM} \sim 60 $ GeV which is a typical wimp mass. Therefore, the linear approximation also
provides a consistent value for the mass of the CDM particles in full agreement 
with CDM simulations.

\medskip

These results show that our theoretical treatment captures many essential
features of dark matter, allowing to determine its nature. When contrasted
to the CDM surface density value obtained from CDM simulations
(instead of the surface density value obtained from observations), 
our approach  gives for the dark matter
particle mass the typical CDM wimps mass scale (GeV), fully consistent
with CDM simulations.
\section{Conclusions}\label{conclu}

Dark matter is characterized by two basic quantities:
the DM particle mass $ m $ and the number of ultrarelativistic
degrees of freedom at decoupling $ g_d $ (or, alternatively  the
decoupling temperature $ T_d $). We obtain the density profiles and theoretical relations
between $ m $ and $ g_d $ involving the observable densities
$ \rho_{DM} $ and  $ \mu_0 $
eqs.(\ref{Qprim}), (\ref{gdu}) and (\ref{muteo}). 
Inserting the observed values of $ \rho_{DM} $ and  $ \mu_0 $ in these
theoretical relations yields $ m, \;  g_d $ and $ Q_p $
eqs. (\ref{dm})-(\ref{mcero}) and (\ref{gdu2}), respectively.

We estimate the galaxy surface density 
and match it with the observed values. Within the same 
scheme, we derive analytically the halo radius $ r_0 $ and the 
factor $ Z $ characterizing 
the reduction of the phase-space density since equilibration till today. 
For these results we use the observed values of the
halo phase-space density $ Q_{halo} $.

\medskip

From the observed values of the surface density we present here clear evidence
that the mass of the DM particle is about one or two keV. 
Evidence based on the phase space density pointing towards a DM particle mass
in the keV scale was presented in refs. \cite{dm1,dm2}.

\medskip

\begin{figure}
\begin{turn}{-90}
\psfrag{"wim100.dat"}{$ \log_{10} \rho_{lin}(r)_{wimp} $ vs. $ r $ in kpc}
\psfrag{"nfwG.dat"}{NFW profile}
\includegraphics[height=9.cm,width=9.cm]{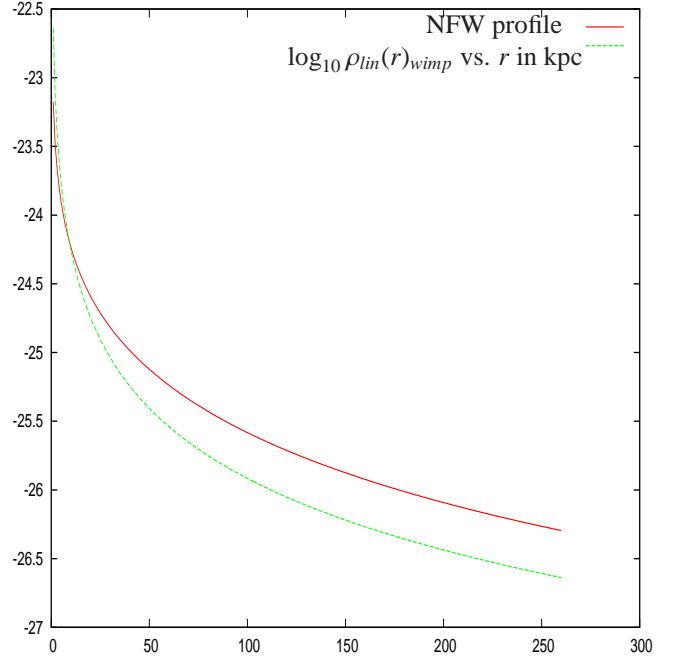}
\end{turn}
\caption{The linear density profile for 100 GeV wimps (broken green line)
and the NFW profile (solid red line) for the same galaxy mass as the 
Burkert profile in fig. \ref{plgg}.
In all cases the densities are in g/${\rm cm}^3 $ and $ r $ in kpc.
The wimps linear density profile follows eq.(\ref{asifin}).
The wimp linear profile exhibits a cusp like the NFW profile.}
\label{perT}
\end{figure}

In addition, one can wonder what would be the results
for heavy wimps. For example, for wimps at $ m_{wimp} = 100 $ GeV 
the characteristic scale $ r_{lin} $ eq.(\ref{rlin1}) takes the value 
given by eq.(\ref{rlinW}). For such {\bf small} $ r_{lin} $ 
the linear profile 
$ \rho_{lin}(r)_{wimp} $ 
appears as a {\bf cusped} profile when observed at scales from 
$ 0.003$ pc to $ 1 $ pc as shown in fig. \ref{perT}. 
Cusped profiles are thus clearly associated 
to heavy DM particles with a huge mass $ m_{wimp} $ 
well above the physical keV scale while cored profiles are 
associated to DM particles with mass in the keV scale. 

\medskip

Notice that the density profile turns out to be cored
or cuspy depending on the DM particle mass $ m $. For $ m \sim $ keV
the resulting density profile is cored as depicted in fig. \ref{plgg}
while for  $ m \gtrsim $ GeV the density profile turns
to be cusped as shown in fig. \ref{perT}.
Figs. \ref{plgg}-\ref{perT} show that the density profiles 
for a 1-2 keV DM particle are 
similar to Burkert (within a factor 2-3, irrelevant 
for the aims of this paper) while for a wimp DM particle, the 
density profile is similar to a NFW profile.

\medskip

Despite its limitations, it is rather remarkable that the linear approximation
is able to reproduce the observations within one order of magnitude.
In the present paper we restrict ourselves to estimate the DM particle mass.
In order to theoretically realize galaxy formation, $N$-body
simulations must be performed with the appropriate primordial power spectrum.
Such spectrum crucially depends for small scales
on the value of the DM particle mass.

\medskip

It must be stressed that the framework presented here
applies to any kind of DM particles: particles with mass in the keV scale
reproduce all observed galaxy magnitudes within one order of magnitude,
while wimps ($ m  \sim 100 $ GeV) present discrepancies with observations
of up to eleven orders of magnitude. This is a robust
indication that the DM particle mass is in the keV scale.

\section*{acknowledgments}

P. S. thanks the Observatoire de Paris-LERMA and the LPTHE
for the kind hospitality extended to him.

\appendix

\section[]{The average phase space-density}\label{averps}

$ Q_{halo} $ in sec. \ref{denesp} follows averaging 
$ \rho(r) $ and $ v^2_{halo}(r) $
over the volume. We define their average using the density 
$ \rho(r) $ eq.(\ref{bur})
as weight function:
\be\label{promQ}
\hskip -1cm {\overline \rho} \equiv 
\frac{\int_0^{R_{vir}} r^2 \; \rho^2(r) \; dr}{\int_0^{R_{vir}} 
r^2 \; \rho(r) \; dr}
\quad , \quad {{\overline {v^2}}_{halo}} \equiv 
\frac{\int_0^{R_{vir}} r^2 \; \rho(r) \; v^2_{halo}(r) \; 
dr}{\int_0^{R_{vir}} r^2 \; \rho(r) \; dr} \; .
\ee
The virial radius $ R_{vir} $ is defined by the radius where the mass computed
from the Burkert profile eq.(\ref{bur}) takes the value \citep{sal07}
\be\label{masemp}
M(R_{vir}) \simeq 10^{12} \; M_{\odot} \; 
\left(\frac{R_{vir}}{259 \; {\rm kpc}}\right)^3 \; .
\ee
Here,
\bea\label{masbur}
&&M(R_{vir}) = 4 \, \pi \int_0^{R_{vir}} r^2 \; \rho(r) \; dr =
 2 \, \pi \; \rho_0 \; r_0^3\cr \cr
&&\times\left[ \ln(1+{\hat c}) -\arctan {\hat c} + \frac12 \; 
\ln(1+{\hat c}^2) \right] \quad ,  \cr \cr
&& {\hat c} \equiv \frac{R_{vir}}{r_0} \; .
\eea
Elliminating $ M(R_{vir}) $ between eqs.(\ref{masemp}) and (\ref{masbur}) gives $ {\hat c} $ 
as a function of $ \rho_0 $ through the trascendental equation
$$
\frac{\rho_0}{0.6187 \; 10^{-27} \; \frac{\rm g}{{\rm cm}^3}} =\frac{{\hat c}^3}{
\ln(1+{\hat c}) -\arctan {\hat c} + \frac12 \; \ln(1+{\hat c}^2)} \; .
$$
The right hand side is a monotonically increasing function of $ {\hat c} $. This
implies that $ {\hat c} $ increases when $ \rho_0 $ increases. Since $ r_0 $ 
decreases when $ \rho_0 $ increases (keeping constant the surface density 
$ \mu_0 $), therefore $ {\hat c} $ increases when $ r_0 $ decreases. 
For the galaxies in Table 1, we find 
$ 9.2 \lesssim {\hat c}\lesssim 24.9 \; , \; 
120 \; {\rm kpc} < R_{vir} < 478 \; {\rm kpc} $,
smaller values of $ {\hat c} $ corresponding to larger galaxies. 

\bigskip

From eqs.(\ref{defQ}), (\ref{defQ}) and (\ref{vbur})
evaluating the integrals in eq.(\ref{promQ}), we find
\bea
&& {\overline \rho} = 0.0662 \; \rho_0 \quad , \quad 
{{\overline {v^2}}_{halo}} = 2.316 \; G \; \rho_0 \; r_0^2 \; , 
\label{vhprom} \quad , \cr \cr
&& Q_{halo} = 3^\frac32 \; 
\frac{\overline \rho}{({\overline v^2_{halo}})^\frac32} =  
\frac{0.069}{G^\frac32 \; \sqrt{\rho_0} \; r_0^3} \; .
\eea

\medskip

For the NFW profile eq.(\ref{nfw}) the virial mass takes the form
$$
M(R_{vir}) = 4 \, \pi \int_0^{R_{vir}} r^2 \; \rho(r) \; dr =
4 \; \pi \; \rho_s \;
r_s^3 \; \left[ \ln(1+c) - \frac{c}{1+c} \right] \quad , 
$$
$$
c \equiv \frac{R_{vir}}{r_s} \; ,
$$
and therefore we find for $ \rho_s $,
\be\label{r0nfw}
\rho_s = 0.310\; 10^{-27} \; 
\frac{\rm g}{{\rm cm}^3} \; \frac{c^3}{\ln(1+c) - 
\displaystyle\frac{c}{1+c}} \; \; .
\ee
The observations give for $ c $ the empirical relation \citep{sal07}
\be\label{cnfw}
c = 9.7 \; \left(\frac{M(R_{vir})}{10^{12} \; M_{\odot}}\right)^{-0.13} \; .
\ee
Therefore, knowing $ M(R_{vir}) $ and $ R_{vir} $ we obtain $ \rho_s $ and 
$ c $ from  eqs.(\ref{r0nfw}) and (\ref{cnfw}).
For the galaxies in Table 1, we find 
$ 23.2 \; {\rm kpc} < r_s < 62.5 \; {\rm kpc} \; ,
\;  0.439 \; 10^{-25} \; {\rm g}/{\rm cm}^3 < \rho_s < 1.087 \; 
10^{-25} \; {\rm g}/{\rm cm}^3
\; , \; 7.64 < c < 13.1 $. We use the values of $ r_s $ and $ \rho_s $ 
for the larger galaxy to plot the NFW curve in fig. \ref{perT}.
Namely, $ r_s = 62.5 $ kpc and $ \rho_s = 1.087 \; 10^{-25} \; 
{\rm g}/{\rm cm}^3 $.

\section[]{The linearized density profile.}\label{perflin}

\begin{figure}
\begin{turn}{-90}
\psfrag{"dengifd.dat"}{$ \Psi_{FD}(y) $}
\psfrag{"bur2.dat"}{$ F_{B}(\alpha \; y) $}
\includegraphics[height=9.cm,width=6.cm]{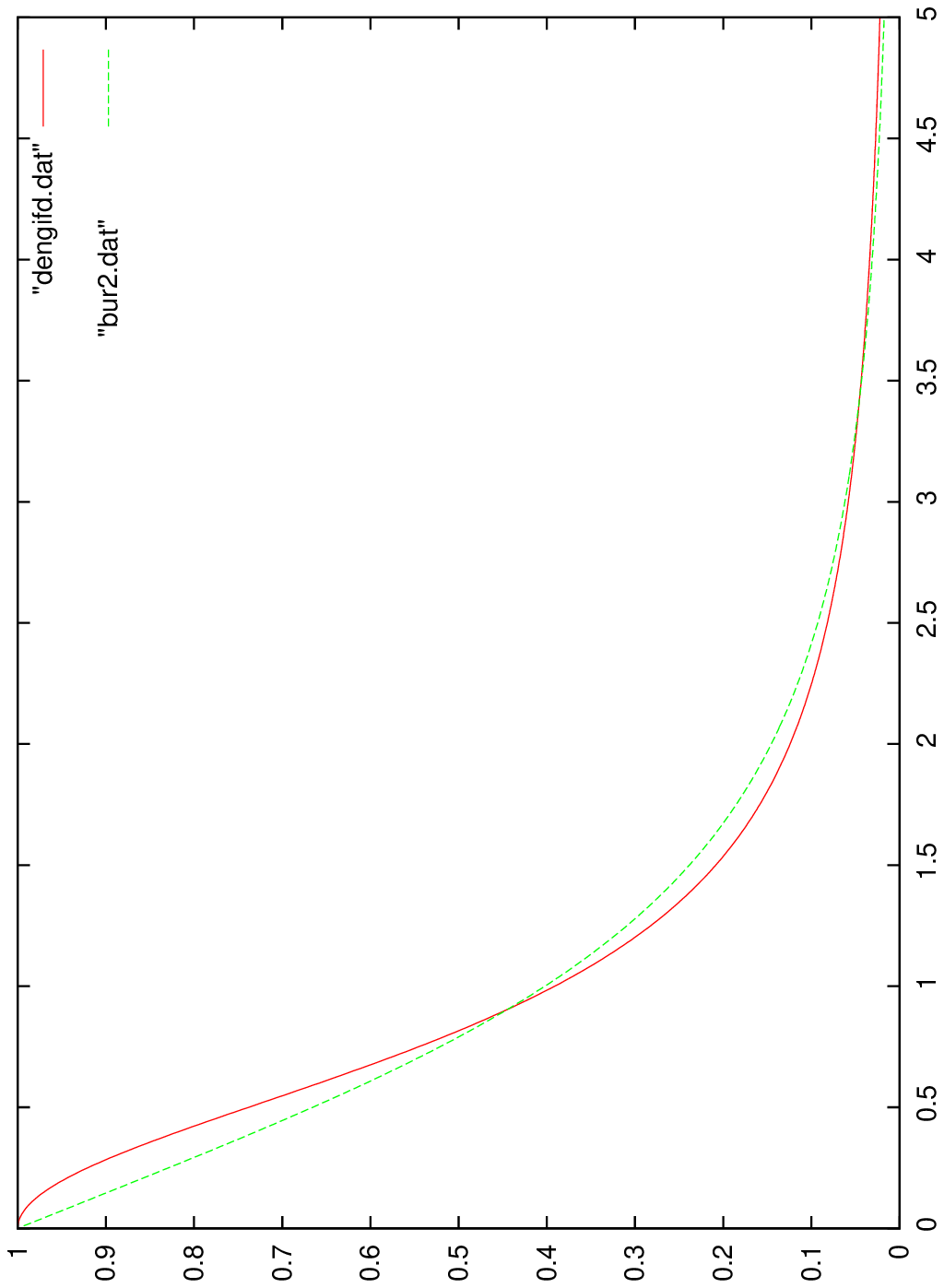}
\psfrag{"dengibe.dat"}{$ \Psi_{BE}(y) $}
\psfrag{"burbe.dat"}{$ F_{B}(\alpha \; y) $}
\includegraphics[height=9.cm,width=6.cm]{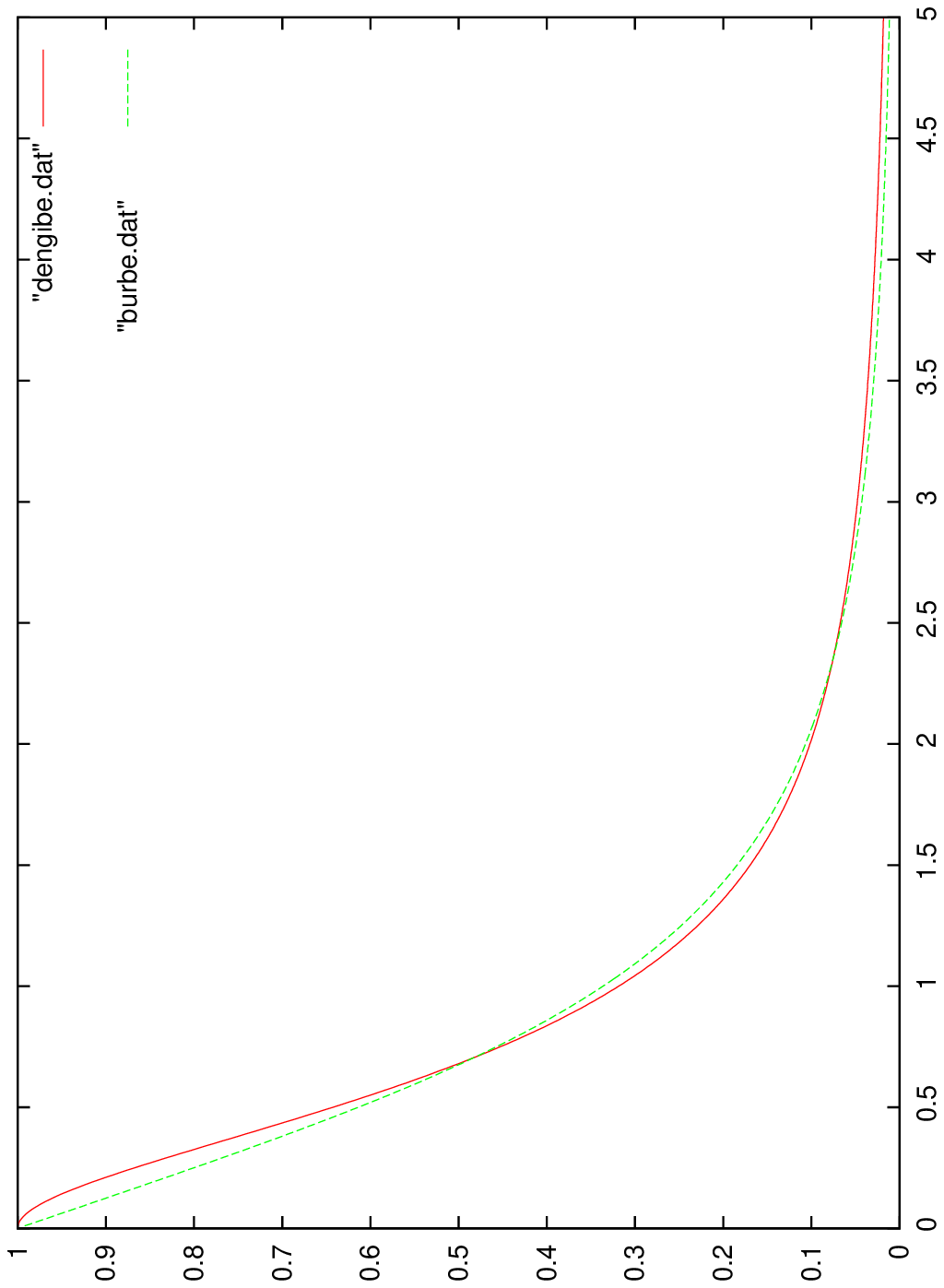}
\psfrag{"dengimb.dat"}{$ \Psi_{MB}(y) $}
\psfrag{"burmb.dat"}{$ F_{B}(\alpha \; y) $}
\includegraphics[height=9.cm,width=6.cm]{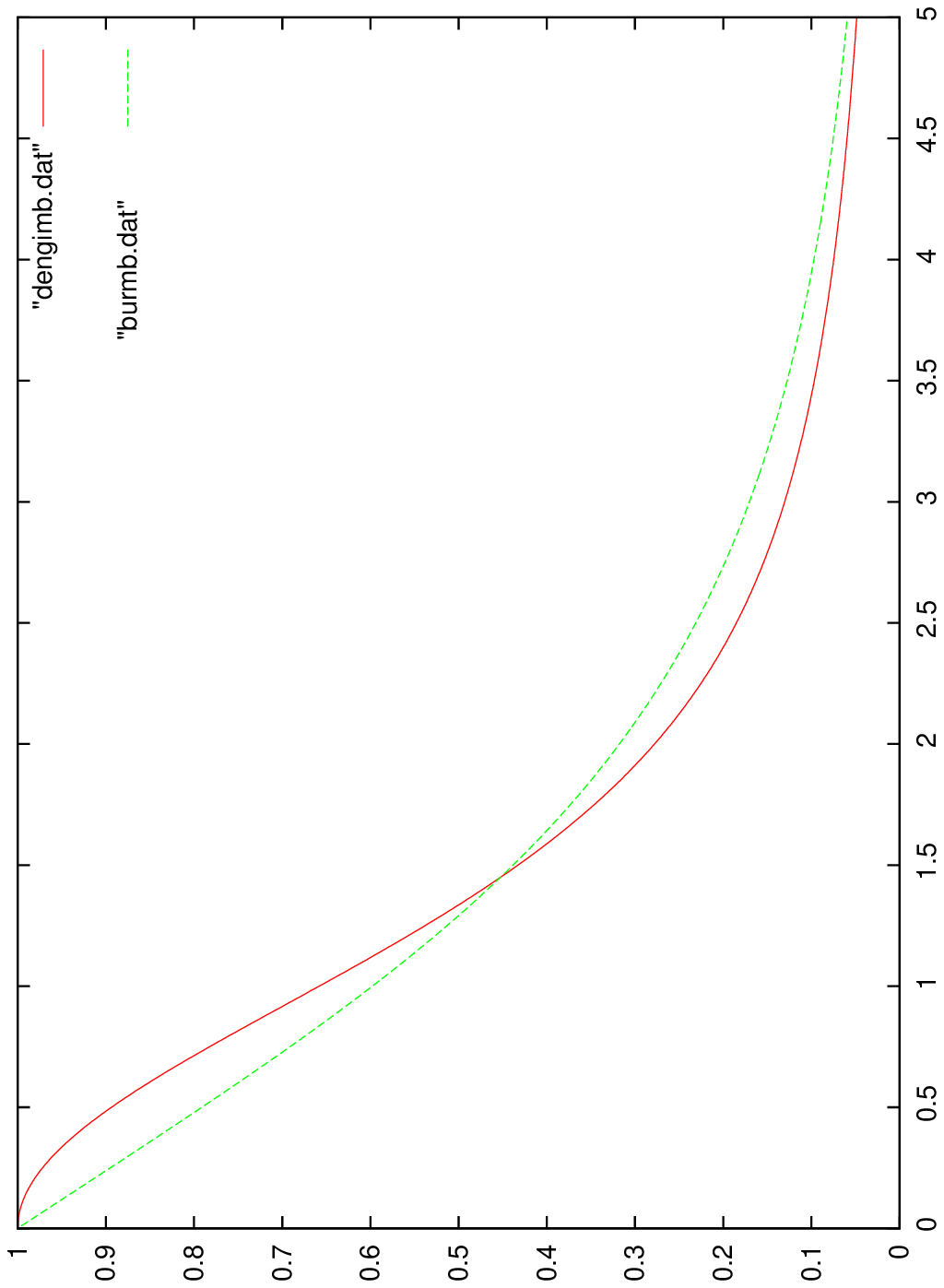}
\end{turn}
\caption{The Burkert profile $ F_{B}(\alpha \; y) $ 
and the linear profile $ \Psi(y) $ computed from first principles
vs. $ y = r/r_{lin} $ for Fermi-Dirac (FD), Bose-Einstein (BE) 
and Maxwell-Boltzmann (MB) statistics. 
The values of $ \alpha $ for each statistics are given in Table B3. 
The linear profile $ \Psi(y) $, especially for
Fermi-Dirac and Bose-Einstein statistics, fits very well
the Burkert profile and as a consequence, $ \Psi(y) $ 
reproduces the observations as well as $ F_{B}(\alpha \; y) $.}
\label{fitpro}
\end{figure}

Both, the Burkert profile $ F_B(r/r_0) $ eq.(\ref{bur}) and the linear 
profile $ \Psi(r/r_{lin}) $ eq.(\ref{perfun}), have the same qualitative 
shape. To make the connection quantitative, 
we fit the linear profile with a Burkert profile setting
\be\label{r0rlin}
x = \alpha \; y \quad , \quad {\rm that ~ is,} \quad 
r_{lin} = \alpha \; r_0 \; .
\ee
We look for the value of $ \alpha $ that gives the best fit by minimizing the
sum of squares:
$$
[ \Psi(y) - F_{B}(\alpha \; y)]^2 \quad {\rm for} \; 0 < y < 3 \; \; .
$$
The best fit for each DM particle statistics is obtained for the values of 
$ \alpha $ reported in Table B3. 
We display in fig. \ref{fitpro} the Burkert profile 
$ F_{B}(\alpha \; y) $ and the 
linear profiles $ \Psi(y) $ for Fermi-Dirac, Bose-Einstein and 
Maxwell-Boltzmann statistics, 
respectively. We see from fig. \ref{fitpro}
that the profiles for Bose-Einstein and Fermi-Dirac statistics are 
better fitted 
by a Burkert profile than the profile for Maxwell-Boltzmann statistics. 

We compute the behaviour of the linear profile $ \rho_{lin}(r) $ 
eq.(\ref{perf}) for 
$ r \gg r_{lin} $ in \ref{asypro}. We find that the linear approximation can be used for
(see \ref{asypro})
$$ 
0 \leq r < r_{max}  \quad {\rm where} \quad  
r_{max} \simeq 8  \; {\rm Mpc} \; .
$$
It must be noticed that the maximum radius $ r_{max} $ turns to be independent
of the DM mass $ m $ and only depends on known cosmological parameters.

We have at the origin $  F_{B}'(0) = - 1 $ while $ \Psi'(0) = 0 $
and  $ \Psi''(0) < 0 $. More precisely $ \Psi''(0) = -2.74 $ for fermionic
DM particles. At the origin, the Burkert profile decreases with unit slope 
while the linear profile has an inverse-parabola shape.

\begin{table*}
 \centering
 \begin{minipage}{140mm}
\begin{tabular}{cc} \hline  
     Particle Statistics & $ \alpha $ \\
\hline 
  Bose-Einstein & $ 0.805 $ \\
\hline 
  Fermi-Dirac & $ 0.688  $ \\
\hline 
  Maxwell-Boltzmann & $ 0.421 $ \\
\hline   
\end{tabular}
\caption{The values of the parameter $ \alpha \equiv r_{lin}/r_0 $ for which
the Burkert profile $ F_{B}(\alpha \; y) $ best fits the
linear profile $ \Psi(y) \equiv \rho_{lin}(r)/\rho_{lin}(0) \; , \; y = r/r_{lin} $.}
\end{minipage}
\end{table*}

Galaxy profiles take an universal form when $ \rho(r)/ \rho_0 $
is expressed as a function of $ r/r_0 $. The Burkert profile is
a particularly simple formula that satisfactorily reproduces the
observations. The linear profile $ \Psi(y) $, especially for
Fermi-Dirac and Bose-Einstein statistics, fits very well
the Burkert profile and therefore, $ \Psi(y) $ is also able
to well reproduce the observations. Namely, the linear profile $ \rho_{lin}(r) $
is well appropriated for small and intermediate scales
$$
0 \leq r < r_{max} \; .
$$
This means that although the linear approximation cannot capture the whole
content of the structure formation, it can well reproduce {\bf universal} features
which are common to all types of galaxies 
as the density profile. Notice that the linear profile $ \Psi(y) $ is universal
as a function of $ y = r/r_{lin} $. The values of $ r_{lin} $ and $ \rho_{lin}(0) $
are not universal and change by orders of magnitude according to the halo mass.
On the contrary, the surface density $ \mu_0 $ defined
by eq.(\ref{denss}) is an universal quantity. Indeed, the theoretical value 
of $ \mu_0 $ that follows from the linear profile $ \rho_{lin}(r) $
eq.(\ref{perf}) can reproduce the observed values of $ \mu_0 $
as it has been shown in \cite{ds1}.

We use this property in section \ref{calcmu} to derive the values of 
the DM particle mass $ m $ and the number of ultrarelativistically degrees
of freedom at decoupling $ g_d $.

\begin{figure}
\begin{turn}{-90}
\psfrag{"dengifd.dat"}{Fermi-Dirac}
\psfrag{"dengibe.dat"}{Bose-Einstein}
\psfrag{"r2dengimb.dat"}{Maxwell-Boltzmann}
\includegraphics[height=9.cm,width=9.cm]{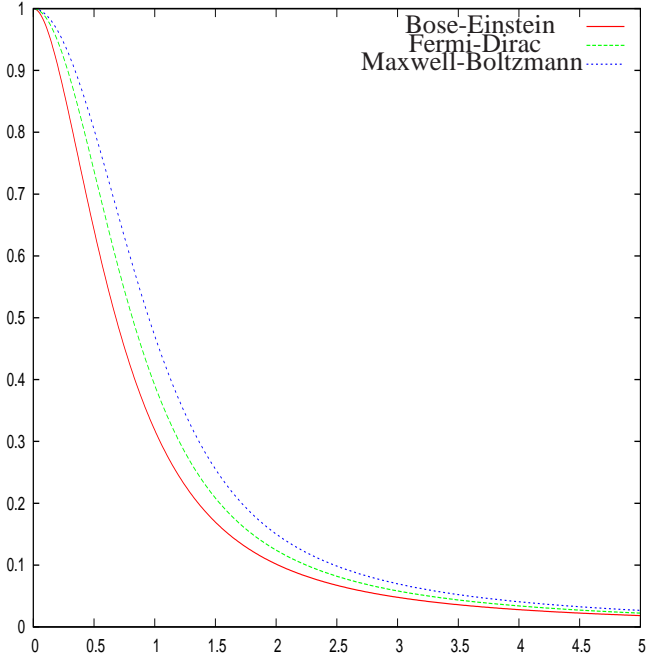}
\end{turn}
\caption{The profiles $ \rho_{lin}(r)/ \rho_{lin}(0) $ vs. $ x $, where $ x \equiv r/r_{lin} $
for Fermions and Bosons decoupling ultrarelativistically and for particles decoupling 
non-relativistically (Maxwell-Boltzmann statistics). The bosons profile is the more peaked,
the MB profile is the shallowest and the fermions profile is lying in-between.
The profiles show little variation with the statistics of the DM particles.}
\label{3perf}
\end{figure}

As shown above the linear profile and the Burkert profile
are the closest for  $ r_{lin} = \alpha \; r_0 $ with $ \alpha = 0.688 $.
On the other hand, we know that the linear approximation always gives values for 
$ r_0 $ larger than the observed values, namely, the linear approximation 
improves for large galaxies \cite{ds1}.
Therefore, we require that $ r_{lin} $ tends to $ r_0 \equiv 0.688 \; r_0 $ 
for large galaxies
which fixes $ b_1 $ to be $ b_1 \simeq 0.8 $. In any case
the dependence of the results on $ b_1 $  [which must be anyway 
$ b_1 \sim 1 $] is quite mild.

\section[]{Asymptotic behaviour of the linear density profile.}\label{asypro}

To derive the asymptotic behaviour of $ \rho_{lin}(r) $ it is convenient 
to change the integration variable in eq.(\ref{perf}) to 
\be\label{vrbls}
\eta \equiv \gamma \; \frac{r}{r_{lin}} \quad , \quad
y = \frac{r}{r_{lin}} \; ,
\ee
and we obtain
\bea\label{rhoasi}
&&\Psi(y) = \frac{\rho_{lin}(r)}{\rho_{lin}(0)} 
= \frac1{y^2 \; \int_0^{\infty} \; \gamma \; N(\gamma) \; d\gamma} \cr \cr
&& \times \int_0^{\infty} N\left(\frac{\eta}{y}\right)
 \; \sin \eta  \; d\eta
\eea
In the limit $ y = r/r_{lin} \to \infty $ we have from eq.(\ref{Nq})
$$
N\left(\frac{\eta}{y}\right) \buildrel{y \gg 1}\over= 
\left(\frac{\eta}{y}\right)^{\frac{n_s}2-1} \; \left[
\ln\left(\frac{c_0}{y} \; q_p^\frac13  \right) + \ln \eta \right]
$$
where we used that $ T(0) = 1 $.

Therefore eq.(\ref{rhoasi}) gives
\bea\label{psi1}
&& \Psi(y)\buildrel{y \gg 1}\over=
\frac{\Gamma\left(\frac{n_s}2\right) \; 
\sin\left(\frac{\pi}4 \; n_s\right)}{\int_0^{\infty} \; 
\gamma \; N(\gamma) \; d\gamma} \; y^{-1-\frac{n_s}2}  \cr \cr
&& \times\left[ 
\ln\left(\frac{c_0}{y} \; q_p^\frac13  \right) + \psi\left(\frac{n_s}2\right) +
\frac{\pi}2 \; \cot \left(\frac{\pi}4 \; n_s \right)\right] \; ,
\eea
where we used the formulas \cite{grad}
\bea
&&\int_0^{\infty} \eta^{\frac{n_s}2-1} \; \sin \eta  \; d\eta = 
\Gamma\left(\frac{n_s}2\right) \; \sin\left(\frac{\pi}4 \; n_s\right) \; ,
\cr \cr\cr
&&\int_0^{\infty} \eta^{\frac{n_s}2-1} \; \sin \eta  \; \ln\eta \; d\eta = 
\Gamma\left(\frac{n_s}2\right) \; \sin\left(\frac{\pi}4 \; n_s\right) \cr\cr
&& \times \left[\psi\left(\frac{n_s}2\right) +
\frac{\pi}2 \; {\rm cotg} \left(\frac{\pi}4 \; n_s \right)\right] \; ,
\eea
$ \psi(x) $ stands for the digamma function.

The asymptotic behaviour eq.(\ref{psi1}) is hence governed by the small $ k $
behaviour of the fluctuations $ \Delta(k,z_{eq}) $ by the end of the radiation
dominated era [see eq.(\ref{flueq})].

Using the numerical values for $ n_s $ and $ c_0 $ 
from eqs.(\ref{val2}) and (\ref{Nq}) and the 
integral over $ N(\gamma) $ eq.(\ref{intNc}), eq.(\ref{psi1}) becomes
\be\label{perasi}
\Psi(y)\buildrel{y \gtrsim  1}\over= \frac{0.4120}{y^{1.482}} \; \frac{ 1
+ 0.1687 \ln\left(\displaystyle \frac{q_p^\frac13}{y}\right)}{1 + 0.04891 \; \ln q} \; . 
\ee
We obtain for DM particles decoupling ultrarelativistically at thermal equilibrium
using eqs.(\ref{dm}) and (\ref{mcero}),
\bea
&&\Psi(y)=0.7705 \; \left(\frac{77.23 \; {\rm kpc}}{r}\right)^{1.482} \; 
\left(\frac{\rm keV}{m}\right)^{1.976} \cr \cr
&& \times \frac{ 1 + 0.1114\ln\left(\displaystyle \frac{\rm kpc}{r}\right)}{1+ 0.2416 \; 
\ln\left(\displaystyle \frac{m}{\rm keV}\right)} \; ,
\eea
where we used $ 1 + n_s/2 = 1.482, \;  2 \; ( 2 + n_s)/3 = 1.976 $.

\begin{figure}
\begin{turn}{-90}
\psfrag{"dengifd.dat"}{Numerical calculation of $ \Psi(y) $ vs. $ y $}
\psfrag{"asifd.dat"}{Asymptotic formula for $ \Psi(y) $ vs. $ y $}
\includegraphics[height=9.cm,width=9.cm]{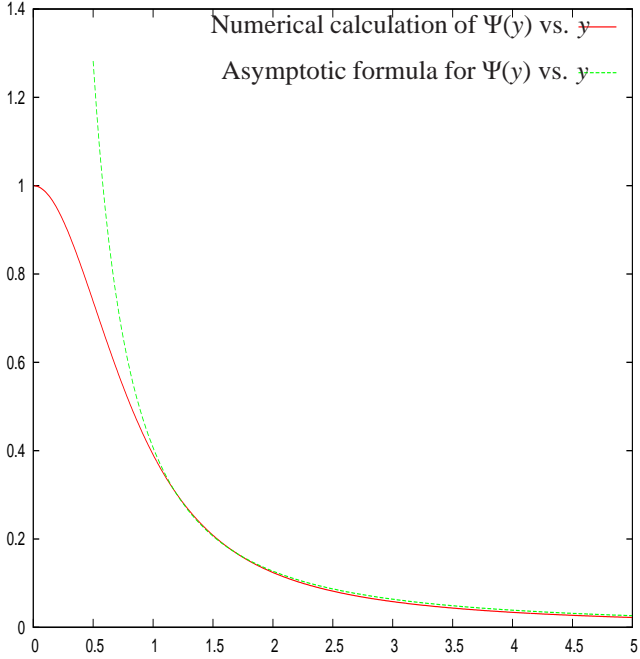}
\end{turn}
\caption{The linear profile $ \Psi(y) $ vs. $ y $ computed from the
numerical Fourier transform eq.(\ref{perfun}) in red continuous line
and computed from the asymptotic formula eq.(\ref{perasi}).
We see that the asymptotic formula well reproduces the linear profile for
$ y \gtrsim 1 $ and not just for $ y \gg 1 $.}
\label{compa}
\end{figure}

We plot in fig. \ref{compa} the asymptotic formula eq.(\ref{perasi}) and the
numerical Fourier transform eq.(\ref{perfun}) for $ \Psi(y) $. We see that the asymptotic
formula eq.(\ref{perasi}) correctly reproduces $ \Psi(y) $ not only for $ y \gg 1 $
but for all $ y \gtrsim 1 $.

We see that there exists a maximum value $ y_{max} $ (and therefore $ r_{max} $)
where the linear profile vanishes:
\be\label{yrmax}
y_{max} = 102.7 \; \left(\frac{m}{\rm keV}\right)^\frac43 \quad , \quad
r_{max} = 7.932 \; {\rm Mpc} \; .
\ee
where we used eqs.(\ref{vrbls}), (\ref{perasi}) and (\ref{qrlin}). 

Notice that $ r_{max} $ turns to be independent
of the DM mass $ m $ and only depends on known cosmological parameters.

Thus, the linear approximation can be used for
$$
0 \leq y < y_{max} \quad , \quad 0 \leq r < r_{max}
$$
where $ \Psi(y) > 0 $ with $ y_{max} $ and $ r_{max} $ given by eq.(\ref{yrmax}).

The nonvalidity  of the linear approximation beyond 8 Mpc reflects the fact
that non-linear effects are important for small wavenumbers: this is
consistent with the fact that we have effectively cutted off the modes 
$ k < k_{eq} $ in the linear approximation [see eq.(\ref{V}) and \cite{ds1}]
as it must be.

Combining the value of $ \rho_{lin}(0) $ in eqs.(\ref{coc1}) and  (\ref{coc2})
with the asymptotic behaviour  eq.(\ref{perasi}) yields
\bea\label{asifin}
\rho_{lin}(r\gtrsim r_{lin})
= 10^{-26} \;  \frac{\rm g}{{\rm cm}^3} \;
\left(\frac{42.03 \; {\rm kpc}}{r}\right)^{1.482} \; 
\ln\left(\frac{7.932\; {\rm Mpc}}{r}\right) && \cr \cr
\times \left[ 1 + 0.04891 \;  \ln q_p\right] \; . \qquad &&
\eea
We then find for DM particles decoupling ultrarelativistically at thermal equilibrium
using eqs.(\ref{dm}) and (\ref{mcero}),
\bea\label{asidur}
\rho_{lin}(r\gtrsim r_{lin})
=10^{-26} \;  \frac{\rm g}{{\rm cm}^3} \;
\left(\frac{36.45 \; {\rm kpc}}{r}\right)^{1.482} \; 
\ln\left(\frac{7.932\; {\rm Mpc}}{r}\right) && \cr \cr
\times  \left[ 1 + 0.2416 \; \ln \left(\frac{m}{\rm keV}\right)\right] \; , \qquad &&
\eea
where $ r_{lin} $ is given by eq.(\ref{qrlin}). It should be remarked
that this behaviour has only a mild logarithmmic dependence on the DM particle
mass $ m $. The scales in eqs.(\ref{asifin})-(\ref{asidur}) only depend on 
known cosmological parameters and not on $ m $.

\medskip

As noticed in \citet{ds1}, the asymptotic decrease of the linear profile
given by eq.(\ref{asifin}) is in remarquable agreement with the universal 
empirical behaviour put forward 
from observations in \cite{wal} and from $\Lambda$CDM simulations in \cite{vass}.
For larger scales we would expect that the contribution from small $ k $ 
modes where nonlinear effects are dominant will give the customary 
$ r^{-3} $ tail exhibited by the Burkert profile eq. (\ref{bur}).

\end{document}